\documentclass[10pt,twocolumn,twoside]{IEEEtran}
\usepackage{url}
\usepackage[cmex10]{amsmath}
\usepackage{amsfonts}
\usepackage{amssymb}
\usepackage{pgf}
\usepackage{graphicx}
\usepackage{grffile}
\usepackage{subfig}
\usepackage[font=scriptsize]{caption}
\usepackage{epstopdf}
\usepackage{mathtools}
\usepackage{tabu}
\usepackage{cite}
\newcommand{\ud}{\,\mathrm{d}}
\ifCLASSINFOpdf
\else
\fi
\begin{document}
%
\title{Some studies on multidimensional Fourier theory for Hilbert transform, analytic signal and space-time series analysis}
%
%
%
\author{Pushpendra Singh$^{*,1,2}$, and Shiv Dutt Joshi$^1$
        \\ $^1$Indian Institute of Technology Delhi, India
\thanks{$^*$Corresponding author's e-mail address: spushp@gmail.com; pushpendra.singh@ee.iitd.ernet.in}  
\thanks{$^2$Jaypee Institute of Information Technology - Noida, India}}
\markboth{Signal/Data analysis \& processing}
{Singh P. \MakeLowercase{\textit{et al.}}: ND-FT, ND-HT.}
\maketitle
\begin{abstract}
In this paper, we propose the Fourier frequency vector (FFV), inherently, associated with multidimensional Fourier transform. With the help of FFV, we are able to provide physical meaning of so called negative frequencies in multidimensional Fourier transform (MDFT), which in turn provide multidimensional spatial and space-time series analysis. The complex exponential representation of sinusoidal function always yields two frequencies, negative frequency corresponding to positive frequency and vice versa, in the multidimensional Fourier spectrum. Thus, using the MDFT, we propose multidimensional Hilbert transform (MDHT) and associated multidimensional analytic signal (MDAS) with following properties: (a) the extra and redundant positive, negative, or both frequencies, introduced due to complex exponential representation of multidimensional Fourier spectrum, are suppressed, (b) real part of MDAS is original signal, (c) real and imaginary part of MDAS are orthogonal, and (d) the magnitude envelope of a original signal is obtained as the magnitude of its associated MDAS, which is the instantaneous amplitude of the MDAS. The proposed MDHT and associated DMAS are generalization of the 1D HT and AS, respectively. We also provide the decomposition of an image into the AM-FM image model by the Fourier method and obtain explicit expression for the analytic image computation by 2D-DFT.
\end{abstract}
%
\IEEEpeerreviewmaketitle
\section{Introduction}
The one-dimensional (1D) Hilbert Transform (HT) and analytic signal (AS)~\cite{NDFHT1} have been used extensively in signal processing and information theory since their introduction. In this study, we use the Fourier theory to obtain the multidimensional HT (MDHT) and multidimensional AS (MDAS), as a generalization of 1D cases, respectively.

Let $f(x)$ be a periodic function of time (or space), i.e. $f(x)=f(x+T_x)$, $\forall x\in\mathbb{R}$, then the Fourier series (FS) of $f(x)$ is defined as
\begin{equation}
f(x)=\sum_{k=0}^{\infty}[a_k\cos(k\omega_1 x)+b_k\sin(k\omega_1 x)]  \label{1ddft1}
\end{equation}
 where $\omega_1=\frac{2\pi}{T_x}$, the coefficients $a_{0}$, $a_{k}$ and $b_{k}$ can be obtained by
\begin{equation}
\begin{aligned}
        a_{0}&=\frac{1}{T_x} \int_{-\frac{T_x}{2}}^{\frac{T_x}{2}} f(x) \ud x ,\\
        a_{k}&=\frac{2}{T_x} \int_{-\frac{T_x}{2}}^{\frac{T_x}{2}} f(x) \cos(k\omega_1x) \ud x , \qquad k\ge1\\
        b_{k}&=\frac{2}{T_x} \int_{-\frac{T_x}{2}}^{\frac{T_x}{2}} f(x) \sin(k\omega_1x) \ud x. \qquad k\ge0.
       \end{aligned}
        \label{1ddft2}
\end{equation}
The Fourier amplitude spectrum for a real function is given by $|F_0|=|a_0|$, $|F_k|=\sqrt{a^2_k+b^2_k}$ and phase spectrum by $\angle F_k=\phi_k=\tan^{-1}(-b_k/a_k)$, which implies that $F_k$ can be
written as
\begin{equation}
F_0=a_0, \quad F_k=a_k-jb_k= \frac{2}{T_x} \int_{-\frac{T_x}{2}}^{\frac{T_x}{2}} f(x) e^{(-jk\omega_1x)} \ud x, \label{1ddft2_1}
\end{equation}
i.e. $F_k= |F_k|e^{j\phi_k}\Leftrightarrow a_k=|F_k|\cos(\phi_k), b_k=|F_k|\sin(\phi_k)$ and hence \eqref{1ddft1} can be written as $f(x)=\sum_{k=0}^{\infty}|F_k|\cos(k\omega_1 x-\phi_k)$.

In the Euler's formula $e^{j\phi}=\cos(\phi)+j\sin(\phi)$, the Hilbert showed that (with $\phi=\omega t$) the function $\sin(\phi)$ is the HT of function $\cos(\phi)$. This yields the $\pm\frac{\pi}{2}$ degree phase-shift operator, which is a basic property of the HT. From the Euler's formula, it is easy to show that $\cos(\phi)=\frac{1}{2}[e^{j\phi}+e^{-j\phi}]$ and $\sin(\phi)=\frac{1}{2j}[e^{j\phi}-e^{-j\phi}]$. Using these results, equation~\eqref{1ddft1} can be written as
\begin{equation}
    \begin{aligned}
    f(x)&=\sum_{k=0}^{\infty}[c_k\exp(jk\omega_1 x)+c^*_k\exp(-jk\omega_1 x)] \Leftrightarrow \\
    f(x)&=\sum_{k=-\infty}^{\infty}[c_k\exp(jk\omega_1 x)]
    \end{aligned}
    \label{1ddft3}
\end{equation}
where $c_0=a_0$, $c_k=\frac{(a_k-jb_k)}{2}$ for $k\ne0$, from \eqref{1ddft2} and \eqref{1ddft3} $c_k$ can be written as
\begin{equation}
c_k=\frac{1}{T_x} \int_{-\frac{T_x}{2}}^{\frac{T_x}{2}} f(x) \exp(-jk\omega_1 x) \ud x. \label{1ddftt4}
\end{equation}
From \eqref{1ddft3}, it is clear that the extra term, $\exp(-jk\omega_1 x)$, which here corresponds to negative frequency, is introduced only due to complex exponential representation of sinusoids in \eqref{1ddft1}, otherwise this extra and redundant term is not required.

We consider the HT, for the sinusoidal functions, as the $\frac{\pi}{2}$ degree phase delay operator, i.e.
\begin{equation}
    \begin{aligned}
    H\{\cos(\phi)\}&=\cos(\phi-\frac{\pi}{2})=\sin(\phi), \\
    H\{\sin(\phi)\}&=\sin(\phi-\frac{\pi}{2})=-\cos(\phi),
    \end{aligned}
    \label{1ddft4_1}
\end{equation} 
where $\phi=\omega_1x$ for 1D, $\phi=\omega_1x+\omega_2y$ for 2D and $\phi=\omega_1x+\cdots+\omega_Mx_M$ for MD function, the HT of function $f(x)$ is denoted as $H\{f(x)\}=\hat{f}(x)$. The HT of constant ($a_0$) is zero, and it can be easily verified by \eqref{1ddft4_1} as $H\{a_0\cos(0)\}=a_0\sin(0)=0$ and $H\{a_0\sin(\pi/2)\}=-a_0\cos(\pi/2)=0$. The HT of complex sinusoidal (complex exponential) signal is multiplication of $-j$ with signal itself (e.g. $H\{\exp(j\omega t)\}=\exp(j[\omega t-\pi/2])=-j\exp(j\omega t)$). By using \eqref{1ddft4_1}, we evaluate the Hilbert transform (HT) of \eqref{1ddft1} and obtain
\begin{equation}
\hat{f}(x)=\sum_{k=0}^{\infty}[a_k\sin(k\omega_1 x)-b_k\cos(k\omega_1 x)].  \label{1ddft5}
\end{equation}
From \eqref{1ddft1} and \eqref{1ddft5} or simply from \eqref{1ddft3}, the analytic signal (AS) is defined as \cite{NDFHT0}
\begin{equation}
    \begin{aligned}
    z(x)&=2\sum_{k=0}^{\infty} [c_k\exp(jk\omega_1 x)], \\
    z(x)&=f(x)+j\hat{f}(x)=r(x) e^{j\phi(x)}
    \end{aligned}
    \label{1ddft6}
\end{equation}
where $r(x) =[f^2(x)+\hat{f}^2(x)]^{\frac{1}{2}}$ is a magnitude envelope of a signal $f(x)$, $\phi(x)=\tan^{-1}[\frac{\hat{f}(x)}{f(x)}]$, real part of AS is original signal and imaginary part is the Hilbert transform of original signal.

\textbf{Observation:} We observe that for real periodic function $f(x)$
\begin{equation}
\sum_{k=-\infty}^{\infty}[a_k\sin(k\omega_1 x)-b_k\cos(k\omega_1 x)]=0.  \label{1ddft7}
\end{equation}
and this can be easily proved from \eqref{1ddft3} by equating imaginary part to zero. Notice the difference between \eqref{1ddft5} and \eqref{1ddft7}, in the range of values of $k$ in summation.

Notice that this definition of AS satisfies the following properties: (a) the extra and redundant negative frequencies, introduced due to complex exponential representation, are suppressed i.e. $k$ takes only positive values in \eqref{1ddft6}, (b) real part of analytic signal is original signal, (c) real and imaginary part of AS are orthogonal, and (d) the magnitude envelope of a real signal is obtained as the magnitude of its associated AS, which is the instantaneous amplitude of the AS.

The frequency, in Hertz (Hz), is defined as number of events or cycles per second. The fundamental period, $T_x$, is the duration of one cycle, and is the reciprocal of the fundamental frequency $f_1$, i.e. $f_1=\frac{1}{T_x}$. Hence, by definition, the frequency is positive physically. In real world, the negative frequency does not exist and the meaning of negative frequencies is only mathematical, and not physical.

The extension of the HT and AS to the two-dimensional (2D) case and their applications to image processing has been limited, due to the non-uniqueness of the
MDHT and MDAS. This fact has led to a variety of definitions with different approaches, to satisfy the 1D conditions in the 2D case, such as the conventional 2D HT in the spatial domain~\cite{NDFHT2} where the negative frequencies are not suppressed, directional HT and quaternionic 2D-AS \cite{NDFHT7}, single orthant 2D-HT \cite{NDFHT8} where real part of analytic signal is not original signal, monogenic signal that is based on the Riesz transform instead of the HT \cite{NDFHT9}, generalized radial HT \cite{NDFHT6}, 2D-HT and its corresponding AS based on a combination of 1D-HTs \cite{NDFHT10}. These approaches have some
useful applications, however the 2D-AS obtained with them do not satisfy the all properties of 1D AS.
There are many interesting applications of the 2D-HT and the corresponding AS, such as edge detection \cite{NDFHT6}, corner detection \cite{NDFHT2}, phase congruency calculations \cite{NDFHT5} and AM-FM image models \cite{NDFHT3,NDFHT4} including others.

The fundamental property of the analytic signal, especially from the viewpoint of image processing and recognition, is the split of identity~\cite{NDFHT9}. The complex AS in polar representation yields two local features, the instantaneous amplitude and instantaneous phase. These local features fulfill the property of invariance and equivariance~\cite{GGKH}, i.e. the local phase depends only on the local structure, and the local amplitude depends only on the local energy (square of amplitude). If these local features are a complete description of a signal, they are said to perform a split of identity~\cite{NDFHT11}. The split of identity is valid only for band-limited signals with local zero mean property~\cite{NDFHT9}. Thus, the split of identity is valid for all the sinusoidal functions and, hence, valid for the Fourier representation of a function. Therefore, the Fourier theory yields the AS representation of a signal that relies on an orthogonal decomposition of the structural information (local phase), and the energetic information (local amplitude).

In this study, we present the multidimensional trigonometric Fourier series, generalize AS representation and HT using the Fourier theory for multidimensional function with following properties: (P1) In the Fourier spectrum of MDAS, the extra and redundant frequencies should be suppressed. (P2) Real part of MDAS is original signal. (P3) Real and imaginary part of MDAS are orthogonal, and (P4) the magnitude envelope of a MD real signal is obtained as the magnitude of its associated MDAS, which is the instantaneous amplitude of the MDAS.

In the literature, there are various methods and applications \cite{rs1,rs13,rs14,rs19,TaylorsNP,LINOEP,PSEEG} of 1D nonlinear and nonstationary time series. Recently, based on the Fourier theory, the Fourier decomposition method (FDM) for nonlinear and nonstationary time series analysis is proposed in \cite{NDFHT0}.  We, in this study, present an extension of the FDM for (2D data) image signal, refer as 2D-FDM, that yields multi-component AM-FM image model.

 This paper is organized as follows: the multidimensional Fourier series, Hilbert transform and analytic signal are discussed in Section 2. The 2D-DTFT and associated analytic signal is discussed in Section 3. The 2D-FDM for AM-FM image model is discussed in Section 4. Section 5 introduces the single orthant MD-DTFT. Numerical results are given in Section 6. Section 7 presents conclusions.
\section{The multidimensional Fourier Series, Hilbert transform and Analytic signal}
In this section, we discuss the 2D Fourier Series (2D-FS), MD-FS, introduce the concept of Fourier frequency vector (FFV) and obtain associated analytic signal.
Let $f(x,y)$ be a periodic (with period $T_x, T_y$) function of 2D space (or 1D space-time)  i.e. $f(x+T_x,y)=f(x,y+T_y)=f(x,y)$, $\forall x,y\in \mathbb{R}$, then the 2D Fourier series (2D-FS) of $f(x,y)$ can be defined as
\begin{multline}
f(x,y)= \sum_{l=0}^{\infty} \sum_{k=0}^{\infty}\Big[ a_{k,l} \cos(k\omega_1x+l\omega_2y) + \\b_{k,l} \sin(k\omega_1x+l\omega_2y) \Big] + \sum_{l=-\infty}^{-1} \sum_{k=1}^{\infty}\Big[ a_{k,l} \cos(k\omega_1x+l\omega_2y) \\+ b_{k,l} \sin(k\omega_1x+l\omega_2y) \Big], \label{FS2D_1}
\end{multline}
where $\omega_1=\frac{2\pi}{T_x}=2\pi u$, $\omega_2=\frac{2\pi}{T_y}=2\pi v$. Since, $l$ is taking both negative and positive values, therefore individual spatial frequencies $l\omega_2$ can be positive as well as negative, but over all resultant spatial frequency (rad/m) is always positive, which is given by $\omega=\sqrt{(k\omega_1)^2+(l\omega_2)^2}$ and negative sign only helps in determining the direction, $\theta=\tan^{-1}(k\omega_1/l\omega_2)$, of wave. Hence, with respect to Fourier theory, we refer them ($k\omega_1,l\omega_2$) as Fourier frequency vector (FFV) that has a magnitude as well as direction and can be written as $\boldsymbol{\omega}=\begin{bmatrix}k\omega_1 & l\omega_2 \end{bmatrix}^{T}$ (in physics, it is similar to a `wave vector' in multidimensional systems). The coefficients $a_{0,0}$, $a_{k,l}$ and $b_{k,l}$ can be obtained by
\begin{multline}
a_{0,0}=\frac{1}{T_xT_y} \int_{-\frac{T_y}{2}}^{\frac{T_y}{2}}\int_{-\frac{T_x}{2}}^{\frac{T_x}{2}} f(x,y) \ud x \ud y, \\
a_{k,l}=\frac{2}{T_xT_y} \int_{-\frac{T_y}{2}}^{\frac{T_y}{2}}\int_{-\frac{T_x}{2}}^{\frac{T_x}{2}} f(x,y) \cos(k\omega_1x+l\omega_2y) \ud x \ud y, \\
b_{k,l}=\frac{2}{T_xT_y} \int_{-\frac{T_y}{2}}^{\frac{T_y}{2}}\int_{-\frac{T_x}{2}}^{\frac{T_x}{2}} f(x,y) \sin(k\omega_1x+l\omega_2y) \ud x \ud y. \label{FS2D_2}
\end{multline}
The Fourier amplitude spectrum for a real function is given by $|F_{0,0}|=|a_{0,0}|$, $|F_{k,l}|=\sqrt{a^2_{k,l}+b^2_{k,l}}$ and phase spectrum
by $\angle F_{k,l}=\phi_{k,l}=\tan^{-1}(-b_{k,l}/a_{k,l})$, which implies that $F_{k,l}$ can be written as
\begin{multline}
F_{k,l}=\frac{2}{T_xT_y} \int_{-\frac{T_y}{2}}^{\frac{T_y}{2}}\int_{-\frac{T_x}{2}}^{\frac{T_x}{2}} f(x,y) e^{-j(k\omega_1x+l\omega_2y)} \ud x \ud y=\\
a_{k,l}-jb_{k,l}, \qquad F_{0,0}=a_{0,0}.  \label{FS2D_2_1}
\end{multline}
We can also consider \eqref{FS2D_1} with (a) limits of double sum from $k=-\infty \text{ to } \infty$, $l=0 \text{ to } \infty$ and results would be same as there is no change in \eqref{FS2D_2}, (b) limits of double sum from $k=-\infty \text{ to } \infty$, $l=-\infty \text{ to } \infty$ and values of $a_{k,l}$ and $b_{k,l}$ in \eqref{FS2D_2} would get divided by two.

\textbf{Discussion:} There is a need for defining Fourier frequency vector in multidimensional Fourier representations.
To demonstrate this need, let $f(x,y)=\cos(3\omega_1 x-4\omega_2 y)$ be a periodic signal (standing wave in 2D space, like image), with $\omega_1=\frac{2\pi}{T_x}$, $\omega_2=\frac{2\pi}{T_y}$, $T_x=3$, $T_y=4$. From \eqref{FS2D_2}, we find only two solutions, $a_{k,l}=1$ if ($k=3$, $l=-4$) or ($k=-3$, $l=4$); $b_{k,l}=0$ for all $k,l$.
Interestingly, both solutions are same as $\cos(3\omega_1 x-4\omega_2 y)=\cos(-3\omega_1 x+4\omega_2 y)$. In situation like this, we cannot avoid negative values of $l$ (or $k$), because there are no solutions with positive values (i.e. nonnegative integers $\mathbb{N}_0=\{0,1,2,\cdots\}$) of $k$ and $l$, which represent Fourier frequencies. This kind of situation does not arise in 1D Fourier representation as $k$ takes values only zero onward in \eqref{1ddft1}. Thus, for the multidimensional Fourier representation, we define the concept of FFV and in this case FFV is $\boldsymbol{\omega}=\begin{bmatrix}3\omega_1 & -4\omega_2\end{bmatrix}^{T}$, which has a magnitude and direction, like any other vector. The magnitude of FFV is frequency $\omega=|\boldsymbol{\omega}|=\sqrt{(3\omega_1)^2+(4\omega_2)^2}$, which is always positive by definition itself. The complex exponential representation of this function, $\cos(3\omega_1 x-4\omega_2 y)=e^{j(3\omega_1 x-4\omega_2 y)}+e^{-j(3\omega_1 x-4\omega_2 y)}$, introduces the extra frequencies by second term $e^{-j(3\omega_1 x-4\omega_2 y)}$.

\textbf{Observation:} We can also use \eqref{FS2D_1} for space-time $(x,t)$ series analysis, e.g. 1D wave equation, $f(x,t)=\cos(k\omega_1 t-l\omega_2 x)$, where $k\omega_1=\frac{2\pi k}{T_1}=2\pi kf_1=2\pi f$, wave vector (or FFV) $\boldsymbol{\omega}=l\omega_2$, wave number (or spatial frequency) $|\boldsymbol{\omega}|=\frac{2\pi |l|}{\lambda_2}=\frac{2\pi}{\lambda}$ and phase velocity $v_p=\frac{k\omega_1}{|\boldsymbol{\omega}|}=f\lambda$.

By the Euler's formula, we know that $\cos(\phi)=\frac{1}{2}[e^{j\phi}+e^{-j\phi}]$ and $\sin(\phi)=\frac{1}{2j}[e^{j\phi}-e^{-j\phi}]$, using these values we can write
\eqref{FS2D_1} as
\begin{multline}
f(x,y)= \sum_{l=0}^{\infty} \sum_{k=0}^{\infty}\Big[ \frac{(a_{k,l}-jb_{k,l})}{2} \exp[j(k\omega_1x+l\omega_2y)] \\+ \frac{(a_{k,l}+jb_{k,l})}{2} \exp[-j(k\omega_1x+l\omega_2y)] \Big]\\ + \sum_{l=-\infty}^{-1} \sum_{k=1}^{\infty}\Big[ \frac{(a_{k,l}-jb_{k,l})}{2} \exp[j(k\omega_1x+l\omega_2y)] \\+ \frac{(a_{k,l}+jb_{k,l})}{2} \exp[-j(k\omega_1x+l\omega_2y)] \Big].\label{FS2D_3}
\end{multline}
This equation can be written as
\begin{multline}
f(x,y)= \sum_{l=0}^{\infty} \sum_{k=0}^{\infty} \Big[c_{k,l} \exp[j(k\omega_1x+l\omega_2y)] \\+  c^*_{k,l}\exp[-j(k\omega_1x+l\omega_2y)]\Big] \\+ \sum_{l=-\infty}^{-1} \sum_{k=1}^{\infty} \Big[c_{k,l} \exp[j(k\omega_1x+l\omega_2y)] \\+ c^*_{k,l}\exp[-j(k\omega_1x+l\omega_2y)]\Big] \Leftrightarrow  \\f(x,y)= \sum_{k=-\infty}^{\infty} \sum_{l=-\infty}^{\infty} c_{k,l} \exp[j(k\omega_1x+l\omega_2y)],\label{FS2D_4}
\end{multline}
where $c_{0,0}=a_{0,0}$ and $c_{k,l}=\frac{(a_{k,l}-jb_{k,l})}{2}$ for $k,l =-\infty,\cdots, 0,1,\cdots,\infty$. Hence, from \eqref{FS2D_2} we can write
\begin{multline}
c_{k,l}=\frac{1}{T_xT_y} \int_{-\frac{T_y}{2}}^{\frac{T_y}{2}}\int_{-\frac{T_x}{2}}^{\frac{T_x}{2}} f(x,y) \exp(-j[k\omega_1x+l\omega_2y])\\ \ud x \ud y. \label{FS2D_5}
\end{multline}
From \eqref{FS2D_4}, we observe that the extra term, $\exp[-j(k\omega_1x+l\omega_2y)]$, which may corresponds to positive, negative, or both frequencies, is introduced only due to complex exponential representation of sinusoids in \eqref{FS2D_1}, otherwise this extra and redundant term is not required.

By using \eqref{1ddft4_1}, we evaluate the multidimensional Hilbert transform (MDHT) of \eqref{FS2D_1} and obtain
\begin{multline}
\hat{f}(x,y)= \sum_{l=0}^{\infty} \sum_{k=0}^{\infty}\Big[ a_{k,l} \sin(k\omega_1x+l\omega_2y) \\- b_{k,l} \cos(k\omega_1x+l\omega_2y) \Big]\\ + \sum_{l=-\infty}^{-1} \sum_{k=1}^{\infty}\Big[ a_{k,l} \sin(k\omega_1x+l\omega_2y) \\- b_{k,l} \cos(k\omega_1x+l\omega_2y) \Big], \label{FS2D_5_1}
\end{multline}
From \eqref{FS2D_1} and \eqref{FS2D_5_1} or simply from \eqref{FS2D_3}, we define multidimensional AS (MDAS)
\begin{equation}
\begin{aligned}
        z(x,y) & =2\sum_{l=0}^{\infty} \sum_{k=0}^{\infty}c_{k,l} \exp[j(k\omega_1x+l\omega_2y)]\\
        & +2\sum_{l=-\infty}^{-1} \sum_{k=1}^{\infty}c_{k,l} \exp[j(k\omega_1x+l\omega_2y)],\\
        z(x,y) &=f(x,y)+j \hat{f}(x,y)=r(x,y)e^{j\phi(x,y)},
       \end{aligned} \label{FS2D_6}
\end{equation}
where $r(x,y) =[f^2(x,y)+\hat{f}^2(x,y)]^{\frac{1}{2}}$ is a magnitude envelope of a signal $f(x,y)$, $\phi(x,y)=\tan^{-1}[\frac{\hat{f}(x,y)}{f(x,y)}]$, real part of MDAS is original signal, imaginary part is the MDHT of original signal and MDAS has only originally present frequencies. Clearly, the complex exponential representation of sinusoidal function always yields two frequencies, negative frequency corresponding to positive frequency and vice versa, in the Fourier spectrum. Hence, MDAS suppress the extra and redundant positive, negative, or both frequencies, introduced due to complex exponential representation of multidimensional Fourier spectrum and satisfy all the MDAS properties P1 to P4.
\textbf{Observation:} We observe that for real periodic function $f(x,y)$
\begin{multline}
\sum_{k=-\infty}^{\infty} \sum_{l=-\infty}^{\infty}\Big[ a_{k,l} \sin(k\omega_1x+l\omega_2y) - b_{k,l} \cos(k\omega_1x+l\omega_2y) \Big]\\=0, \label{FS2D_6_1}
\end{multline}
and this can be easily proved from \eqref{FS2D_4} by equating imaginary part to zero.

The above discussion can be easily extended for MD-FS which can be written as
\begin{multline}
f(x_1,\cdots,x_M)=\sum_{k_M=-\infty}^{\infty} \cdots \sum_{k_{3}=-\infty}^{\infty} \sum_{k_{2}=0}^{\infty} \sum_{k_{1}=0}^{\infty} \big[a_{k_1,\cdots,k_M} \\  \cos(k_1\omega_1x_1+\cdots+k_M\omega_Mx_M)\\+b_{k_1,\cdots,k_M} \sin(k_1\omega_1x_1+\cdots+k_M\omega_Mx_M)\big]\\
+\sum_{k_M=-\infty}^{\infty} \cdots \sum_{k_{3}=-\infty}^{\infty} \sum_{k_{2}=-\infty}^{-1} \sum_{k_{1}=1}^{\infty} \big[a_{k_1,\cdots,k_M} \\  \cos(k_1\omega_1x_1+\cdots+k_M\omega_Mx_M)\\+b_{k_1,\cdots,k_M} \sin(k_1\omega_1x_1+\cdots+k_M\omega_Mx_M)\big],\label{FS2D_7_0}
 \end{multline}
\begin{multline}
 f(x_1,\cdots,x_M)=\sum_{k_M=-\infty}^{\infty} \cdots \sum_{k_{2}=-\infty}^{\infty} \sum_{k_{1}=-\infty}^{\infty} c_{k_1,\cdots,k_M} \\  \exp[j(k_1\omega_1x_1+\cdots+k_M\omega_Mx_M)],\\
c_{k_1,\cdots,k_M}=\frac{1}{T_1\cdots T_M} \int_{-\frac{T_M}{2}}^{\frac{T_M}{2}} \cdots \int_{-\frac{T_1}{2}}^{\frac{T_1}{2}} f(x_1,\cdots,x_M) \\ \times \exp[-j(k_1\omega_1x_1+\cdots+k_M\omega_Mx_M)] \ud x_1 \cdots \ud x_M, \label{FS2D_8}
\end{multline}
where $c_{k_1,\cdots,k_M}=(a_{k_1,\cdots,k_M}-jb_{k_1,\cdots,k_M})/2$.
 We can easily obtain MDHT by replacing $\cos$ with $\sin$ and $\sin$ with $-\cos$ in \eqref{FS2D_7_0} and obtain AS such that real part of AS is original signal, AS has positive resultant frequency $\omega=|\boldsymbol{\omega}|=\sqrt{(k_1\omega_1)^2+\cdots+(k_M\omega_M)^2}$ and its FFV can be written as $\boldsymbol{\omega}=\begin{bmatrix}k_1\omega_1 & \cdots & k_M\omega_M \end{bmatrix}^{T}$. The power of the FT can be realized from the fact that the HT and analytic representation of a signal are, inherently, present in the Fourier representation.


\textbf{Observation:} We can also use \eqref{FS2D_7_0} for $(M-1)$D space-time series $(x_1,\cdots,x_{(M-1)},t)$ analysis, e.g. 2D wave equation, $f(x,y,t)=\cos(k_1\omega_1 t-k_2\omega_2 x-k_3\omega_3 y)$, where $k_1\omega_1=\frac{2\pi k_1}{T_1}=2\pi k_1f_1=2\pi f$, wave vector (or FFV) $\boldsymbol{\omega}=\begin{bmatrix}k_2\omega_2 & k_3\omega_3 \end{bmatrix}^{T}$, wave number (or spatial frequency) $|\boldsymbol{\omega}|=\sqrt{(k_2\omega_2)^2+(k_3\omega_3)^2}=\frac{2\pi}{\lambda}$ and phase velocity $v_p=\frac{k_1\omega_1}{|\boldsymbol{\omega}|}=f\lambda$.

For non-periodic MD signal, $f(x_1,\cdots,x_M)$, the MD Fourier transform (MDFT) and inverse MDFT (IMDFT) are defined as
\begin{multline}
C[f_1,\cdots,f_M]=\int_{-\infty}^{\infty} \cdots \int_{-\infty}^{\infty} f(x_1,\cdots,x_M) \\ \times \exp[-j(\omega_1x_1+\cdots+
\omega_Mx_M)] \ud x_1 \cdots \ud x_M,\\
f(x_1,\cdots,x_M)=\int_{-\infty}^{\infty} \cdots \int_{-\infty}^{\infty} C[f_1,\cdots,f_M] \\  \exp[j(\omega_1x_1+\cdots+\omega_Mx_M)]\ud f_1 \cdots \ud f_M,\label{FS2D_81}
\end{multline}
respectively. We define MDAS, for real signal $f(x_1,\cdots,x_M)$, as
\begin{multline}
z(x_1,\cdots,x_M)=2\int_{-\infty}^{\infty} \cdots \int_{-\infty}^{\infty} \int_{0}^{\infty} C[f_1,\cdots,f_M] \\  \exp[j(\omega_1x_1+\cdots+\omega_Mx_M)]\ud f_1 \cdots \ud f_M,\label{FS2D_82}
\end{multline}
so its real part is original signal and imaginary part is HT of original signal. It is very convenient to take first frequency always positive (may be related to 1D time or space) and rest of the frequencies (related to MD space) may be positive or negative or both, which is also clear from~\eqref{1ddft1}, \eqref{FS2D_1}, \eqref{FS2D_7_0} and \eqref{FS2D_82}, and this convention has been used throughout in this study.

\newtheorem{theorem}{Theorem}[section]
\newtheorem{lemma}[theorem]{Lemma}
\newtheorem{proposition}[theorem]{Proposition}
\newtheorem{corollary}[theorem]{Corollary}
\newtheorem{axiom}[theorem]{Axiom}
From the above discussions (1D equations \eqref{1ddft1} and \eqref{1ddft6}; 2D equations \eqref{FS2D_1} and \eqref{FS2D_6}), we propose the following result:
\begin{proposition}
\label{LeftCosetsDisjoint}
Let $s(\mathbf{x})$ be a real valued signal and $z(\mathbf{x})=s(\mathbf{x})+j\hat{s}(\mathbf{x})$ is the AS representation of $s(\mathbf{x})$. Then real-valued Fourier representation (RVFR) of $s(\mathbf{x})$ and complex-valued Fourier representation (CVFR) of $z(\mathbf{x})$ are same, i.e.
$S_r(\boldsymbol{\omega})=Z(\boldsymbol{\omega})$, where $S_r(\boldsymbol{\omega})$ is RVFR of $s(\mathbf{x})$, $Z(\boldsymbol{\omega})$ is CVFR of $z(\mathbf{x})$, $\mathbf{x}=\begin{bmatrix}x_1 & \cdots & x_M \end{bmatrix}^{T}$ and $\boldsymbol{\omega}=\begin{bmatrix}\omega_1 & \cdots & \omega_M \end{bmatrix}^{T}$.
\end{proposition}

For a 2D periodic signal, the RVFR is given by \eqref{FS2D_1} and \eqref{FS2D_2}, the CVFR is given by \eqref{FS2D_4} and \eqref{FS2D_5}.
\section{The 2D-DTFT and associated analytic signal}
Let $x[m,n]$ be a non-periodic and real function of time, then the 2D discrete time Fourier transform (2D-DTFT) of $x[m,n]$ is defined as
\begin{equation}
X(\omega_1,\omega_2)=\sum_{m=-\infty}^{\infty} \sum_{n=-\infty}^{\infty} x[m,n] \exp(-j[\omega_1 m+\omega_2 n])  \label{FDM_eq21}
\end{equation}
 and 2D inverse discrete time Fourier transform (2D-IDTFT) is defined as
\begin{multline}
x[m,n]=\frac{1}{2\pi}\frac{1}{2\pi}\int_{-\pi}^{\pi}\int_{-\pi}^{\pi} X(\omega_1,\omega_2) \exp(j[\omega_1 m+\omega_2 n])\\ \ud \omega_1 \ud \omega_2\label{FDM_eq22}
\end{multline}
 It is easy to show that,
from Eq.~\eqref{FDM_eq21}, $X(-\omega_1,-\omega_2)=X^*(\omega_1,\omega_2)$, $X(-\omega_1,\omega_2)=X^*(\omega_1,-\omega_2)$. We rewrite Eq.~\eqref{FDM_eq22} as
\begin{multline}
x[m,n]=\\\frac{1}{2\pi}\frac{1}{2\pi}\Big [\int_{0}^{\pi} \int_{0}^{\pi} X(\omega_1,\omega_2) \exp(j[\omega_1 m+\omega_2 n]) \ud \omega_1 \ud \omega_2 \\+ \int_{-\pi}^{0} \int_{0}^{\pi} X(\omega_1,\omega_2) \exp(j[\omega_1 m+\omega_2 n]) \ud \omega_1 \ud \omega_2\\ +\int_{-\pi}^{0} \int_{-\pi}^{0} X(\omega_1,\omega_2) \exp(j[\omega_1 m+\omega_2 n]) \ud \omega_1 \ud \omega_2\\+ \int_{0}^{\pi} \int_{-\pi}^{0} X(\omega_1,\omega_2) \exp(j[\omega_1 m+\omega_2 n]) \ud \omega_1 \ud \omega_2 \Big]. \label{FDM_eq23}
\end{multline}
In this Eq., first term (denoted as $z_1[m,n]$ and $0\le\omega_1\le\pi,0\le\omega_2\le\pi$) is complex conjugate of third term ($z^*_1[m,n]$ and $-\pi\le\omega_1\le0,-\pi\le\omega_2\le0$) and second term (denoted as $z_2[m,n]$ and $-\pi\le\omega_2\le0, 0\le\omega_1\le\pi$) is complex conjugate of fourth term ($z^*_2[m,n]$ and $0\le\omega_2\le\pi, -\pi\le\omega_1\le0$). As $x[m,n]$ is real function, we can write
\begin{equation}
    \begin{aligned}
        x[m,n] &=Re\{z_{14}[m,n]\} =Re\{z_{23}[m,n]\}\\
        &=Re\{z_{12}[m,n]\}=Re\{z_{34}[m,n]\},
       \end{aligned} \label{FDM_eq24}
\end{equation}
where $Re\{.\}$ denote real part of analytic signal (AS)
\begin{equation}
 \begin{aligned}
        z_{14}[m,n] & =2(z_1[m,n]+z_2[m,n])= x[m,n]+j\hat{x}_{14}[m,n],  \\
        z_{23}[m,n] & =2(z^*_1[m,n]+z^*_2[m,n])= x[m,n]+j\hat{x}_{23}[m,n],     \\
        z_{12}[m,n] & =2(z_1[m,n]+z^*_2[m,n])= x[m,n]+j\hat{x}_{12}[m,n],   \\
        z_{34}[m,n] & =2(z^*_1[m,n]+z_2[m,n])= x[m,n]+j\hat{x}_{34}[m,n],
       \end{aligned}
 \label{FDM_eq240}
\end{equation}
where $\hat{x}_{14}[m,n]=-\hat{x}_{23}[m,n]$, $\hat{x}_{12}[m,n]=-\hat{x}_{24}[m,n]$ and subscripts denote the quadrants of the Fourier domain considered in AS representation.
The DTFT of these AS, with their frequency supports, can be written as
\begin{equation}
    \begin{aligned}
        Z_{14}(\omega_1,\omega_2) &, \quad  \omega_1\in[0,\pi], & \omega_2\in[-\pi,\pi], \\
        Z_{23}(\omega_1,\omega_2) &, \quad\omega_1\in[-\pi,0], & \omega_2\in[-\pi,\pi],\\
        Z_{12}(\omega_1,\omega_2) &, \quad\omega_1\in[-\pi,\pi], & \omega_2\in[0,\pi],\\
        Z_{34}(\omega_1,\omega_2) &, \quad\omega_1\in[-\pi,\pi], & \omega_2\in[-\pi,0].
       \end{aligned} \label{FDM_eq240_0}
\end{equation}
Notice that with this definition of AS, the extra negative frequencies $\omega_1$ and $\omega_2$ are suppressed in AS $z_{14}[m,n]$ and $z_{12}[m,n]$, respectively, and this preserve the desired property for the AS. The AS $z_{23}[m,n]$ and $z_{34}[m,n]$ are the complex conjugate of $z_{14}[m,n]$ and $z_{12}[m,n]$, respectively. From the analytic image point of view, all four imaginary part of AS in \eqref{FDM_eq240} yield different images.
To obtain the the original signal we need to know the values of 2D-DTFT in first and second or third and fourth, first and fourth or second and third quadrants, which is also clear from the above discussions with equations \eqref{FDM_eq240} and \eqref{FDM_eq240_0}. It is to be noted that the only AS $z_{14}[m,n]$ coincides with the proposed AS in \eqref{FS2D_6} and hence its imaginary part, $Im\{z_{14}[m,n]\}$, coincides with proposed HT in \eqref{FS2D_5_1}.

\emph{Example: Time-frequency analysis of 2D unit sample sequence.}
The unit sample sequence defined as $\delta [m-m_0,n-n_0]=1$ at $m=m_0,n=n_0$ and zero otherwise. We obtain the analytic representation of $x[m,n]=\delta [m-m_0,n-n_0]\Leftrightarrow X(\omega_1,\omega_2)=\exp(-j[\omega_1 m_0+\omega_2 n_0])$ as
$z_{14}[m,n]$ $=2z_1[m,n]+2z_2[m,n]=[\frac{\sin(\pi (m-m_0))}{\pi (m-m_0)}] [\frac{\sin(\pi (n-n_0))}{\pi (n-n_0)}]+j[\frac{(1-\cos(\pi (m-m_0)))\sin(\pi (n-n_0))}{\pi (m-m_0)\pi (n-n_0)}]$, as given in \eqref{FDM_eq24}, its real part is original signal, i.e. $x[m,n]=[\frac{\sin(\pi (m-m_0))}{\pi (m-m_0)}] [\frac{\sin(\pi (n-n_0))}{\pi (n-n_0)}]=\delta [m-m_0,n-n_0]$. We obtain the phase of $z_{14}$ as $\phi[m,n]=\frac{\pi}{2} (m-m_0)$ and hence $\omega_1[m,n]=\frac{\pi}{2}$, which corresponds to half of the Nyquist frequency, and other frequency is $\omega_2[m,n]=0$. Similarly, we obtain AS
$z_{12}[m,n]=2z_1[m,n]+2z^*_2[m,n]=[\frac{\sin(\pi (m-m_0))}{\pi (m-m_0)}] [\frac{\sin(\pi (n-n_0))}{\pi (n-n_0)}]+j[\frac{\sin(\pi (m-m_0))(1-\cos(\pi (n-n_0)))}{\pi (m-m_0)\pi (n-n_0)}]$, as given in \eqref{FDM_eq24}, its real part is original signal, i.e. $x[m,n]=[\frac{\sin(\pi (m-m_0))}{\pi (m-m_0)}] [\frac{\sin(\pi (n-n_0))}{\pi (n-n_0)}]=\delta [m-m_0,n-n_0]$. We obtain the phase of $z_{12}$ as $\phi[m,n]=\frac{\pi}{2} (n-n_0)$ and hence $\omega_2[m,n]=\frac{\pi}{2}$, which corresponds to half of the Nyquist frequency, and other frequency is $\omega_1[m,n]=0$.

If analytic signal $z[m,n]$ is defined as four times of first quadrant, where $0\le\omega_1\le\pi, 0\le\omega_2\le\pi$, then we can observe that
\begin{equation}
x[m,n]\ne Re\{z[m,n]\}, \label{2ddft_pro5}
\end{equation}
where $Re\{.\}$ denote real part of analytic signal (AS)
\begin{multline}
z[m,n]=\frac{1}{\pi}\frac{1}{\pi}\int_{0}^{\pi}\int_{0}^{\pi} X(\omega_1,\omega_2) \exp(j\omega_1 m+j\omega_2 n)\\ \ud \omega_1 \ud \omega_2. \label{2ddft_pro6}
\end{multline}
Notice that with this definition of AS, the negative frequencies are suppressed but real part of AS is not original signal and this does not provide a phase delay of $-\pi/2$ degree.

\section{The 2D Fourier decomposition method and AM-FM image model}
We propose to use the following 2D Fourier decomposition method (2D-FDM) to obtain multi-component AM-FM image model defined as
\begin{equation}
f(x,y)=\sum_{i=1}^{M}g_i(x,y)+n(x,y),  \label{AMFMImage1}
\end{equation}
where $n(x,y)$ is a noise representing any residue (constant or trend) components, and the $g_i(x,y)=a_i(x,y)\cos(\phi_i(x,y))$ are M monocomponent (zero mean narrow band) nonstationary image signals that represents the 2D Fourier intrinsic band functions (2D-FIBFs) for a image signal.
We define 2D monocomponent signals that follow either of the following conditions
\begin{subequations}
\begin{alignat}{2}
    \phi_x(x,y)=\frac{\partial \phi(x,y)}{\partial x}>0, \quad \forall x, \label{momocomp1}\\
    \phi_y(x,y)=\frac{\partial \phi(x,y)}{\partial y}>0, \quad \forall y, \label{momocomp2}
\end{alignat}
\end{subequations}
or both \eqref{momocomp1} and \eqref{momocomp2}.

\textbf{Example:} \emph{monocomponent 2D signal e.g.} (a) $g(x,y)=\cos(3\omega_1 x-4\omega_2 y)$ which follow \eqref{momocomp1} or \eqref{momocomp2} (b) $g(x,y)=\cos(3\omega_1 x+4\omega_2 y)$ which follow both \eqref{momocomp1} and \eqref{momocomp2}.

We write analytic function $Z_{14}[m,n]$ from Eq.~\eqref{FDM_eq240} as
\begin{multline}
\frac{1}{\pi}\frac{1}{2\pi}\int_{-\pi}^{\pi}\int_{0}^{\pi} X(\omega_1,\omega_2) \exp(j[\omega_1 m+\omega_2 n]) \ud \omega_1 \ud \omega_2=\\\sum_{i=1}^M a_{i}[m,n]\exp(j\phi_{i}[m,n]) \label{FDM_eq251}
\end{multline}
where (with $\omega_{10}=0, \omega_{1M}=\pi$)
\begin{multline}
 a_{i}[m,n]\exp(j\phi_{i}[m,n]) =  \frac{1}{\pi}\frac{1}{2\pi}\int_{-\pi}^{\pi} \int_{\omega_{1(i-1)}}^{\omega_{1i}} X(\omega_1,\omega_2) \\ \exp(j[\omega_1 m+\omega_2 n]) \ud \omega_1 \ud \omega_2, \label{FDM_eq261}
\end{multline}
for $i=1,\cdots,M$.
To obtain minimum number of AFIBFs in low to high frequency scan, for each $i$, start with $\omega_{1(i-1)}$, increase and select the maximum value of $\omega_{1i}$ such that $\omega_{1(i-1)} \le \omega_{1i} \le \pi$ and phase $\phi_{i}[m,n]$ is a monotonically increasing function with respect to (w.r.t.) $\omega_{1i}[m,n]$, i.e.
\begin{equation}
a_{i}[m,n]\ge0, \quad \omega_{1i}[m,n]=(\phi_i[m+1,n]-\phi_i[m,n])\ge0, \label{FDM_eq26v11}
\end{equation}
for all $m, n$.
Similarly, in high to low frequency scan, the lower and upper limits of integration in Eq.~\eqref{FDM_eq261} will change to $\omega_{1i}$ to $\omega_{1(i-1)}$, respectively, with $\omega_{10}=\pi, \omega_{1M}=0$, and we can obtain minimum number of AFIBFs by selecting the minimum value of $\omega_{1i}$ such that $0 \le \omega_{1i}\le \omega_{1(i-1)}$ and Eq.~\eqref{FDM_eq26v11} is satisfied. As we have represented $Z_{14}[m,n]$ by equations \eqref{FDM_eq251},\eqref{FDM_eq261}, \eqref{FDM_eq26v11} and considered the decomposition w.r.t. only $\omega_1$, the same way, we can represent $Z_{12}[m,n]$ and consider decomposition w.r.t. only $\omega_2$.

We proposed multivariate FDM (MFDM), based on zero-phase filtering (ZPF), in \cite{NDFHT0}. Here, we propose to use MFDM for the decomposition of 2D signal (e.g. image) into AM-FM model defined in \eqref{AMFMImage1}.
\section{Single orthant MD-DTFT}
In this section, we consider multidimensional (2D and its extension to general MD) DTFT of real multidimensional signals. This approach evaluate DTFT of real signal in only first orthant and values in rest of the orthant are obtained by simple conjugation defined (for 2D and MD cases) as follows:
\subsection{Single orthant 2D-DTFT and 2D-analytic signal}
Let $x[m,n]$ be a non-periodic and real function of time, then the 2D discrete time Fourier transform (2D-DTFT) of $x[m,n]$, we define as
\begin{equation}
X(\omega_1,\omega_2)=\sum_{m=-\infty}^{\infty} \sum_{n=-\infty}^{\infty} x[m,n] \exp(-i\omega_1 m-j\omega_2 n)  \label{2ddft_pro1}
\end{equation}
such that
\begin{equation}
i.j=-1, \qquad i^2=j^2=-1, \label{2ddft_pro3}
\end{equation}
where $i$ and $j$ are purely imaginary number and represent phase corresponding to frequencies $\omega_1$ and $\omega_2$, respectively. Since, there are two frequencies $\omega_1$ and $\omega_2$ and integration of frequency is phase, hence it is logical and natural to define two phases corresponding to two frequencies. Here, we define two conjugation, bar ($-$) and star ($*$) corresponding to $i$ and $j$, respectively. Hence, we write \eqref{2ddft_pro1} as
\begin{multline}
X(\omega_1,\omega_2)= X_r(\omega_1,\omega_2)+ij X_{ij}(\omega_1,\omega_2) \\+i X_{i}(\omega_1,\omega_2)+j X_{j}(\omega_1,\omega_2) \label{2ddftNewform1}
\end{multline}
where, first term of the right side of this equation is first real part (FRP), second term is real part due to product of two imaginary parts corresponding to $i$ and $j$, third and fourth term are imaginary parts.
From \eqref{2ddft_pro3} and \eqref{2ddftNewform1}, we obtain the following equations for each of the four quadrants
\begin{multline}
\begin{aligned}
X_1(\omega_1,\omega_2)= [X_r(\omega_1,\omega_2)-X_{ij}(\omega_1,\omega_2) ]+\\j [X_{i}(\omega_1,\omega_2)+ X_{j}(\omega_1,\omega_2)],\\
X_2(\omega_1,\omega_2)= [X_r(\omega_1,\omega_2)+X_{ij}(\omega_1,\omega_2) ]+\\j [-X_{i}(\omega_1,\omega_2)+ X_{j}(\omega_1,\omega_2)],\\
X_3(\omega_1,\omega_2)= [X_r(\omega_1,\omega_2)-X_{ij}(\omega_1,\omega_2) ]-\\j [X_{i}(\omega_1,\omega_2)+ X_{j}(\omega_1,\omega_2)],\\
X_4(\omega_1,\omega_2)= [X_r(\omega_1,\omega_2)+X_{ij}(\omega_1,\omega_2) ]-\\j [-X_{i}(\omega_1,\omega_2)+ X_{j}(\omega_1,\omega_2)],
\end{aligned}
 \label{2ddftNewform1_1}
\end{multline}
where third quadrant is complex conjugate of first one, and fourth quadrant is complex conjugate of second one.
It can be shown that the \eqref{FDM_eq21} and \eqref{2ddftNewform1_1} are actually different representation of the same thing.
We now explore the properties of this 2D-DTFT: \\(1) \textbf{Symmetry}: For real $x[m,n]$, it is easy to show, from Eq.~\eqref{2ddft_pro1}, that (a) $\bar{X}(\omega_1,\omega_2)=X(-\omega_1,\omega_2)$, (b) $\bar{X}^*(\omega_1,\omega_2)=X(-\omega_1,-\omega_2)$, (c) ${X}^*(\omega_1,\omega_2)=X(\omega_1,-\omega_2)$. This clearly indicates that, if we know the value of $X(\omega_1,\omega_2)$ in first quadrant ($0\le\omega_1\le\pi$ and $0\le\omega_2\le\pi$) only, then we can obtain the value of $X(\omega_1,\omega_2)$ in all the quadrants. The spectrum ($|{X}(\omega_1,\omega_2)|$) of this 2D-DTFT is same as 2D-DFT defined in \eqref{FDM_eq21} which is Hermitian symmetric.
\\(2) \textbf{Periodicity}: ${X}(\omega_1,\omega_2)$ is periodic in $\omega_1$, $\omega_2$  with period $2\pi$, i.e. ${X}(\omega_1,\omega_2)={X}(\omega_1+2\pi,\omega_2+2\pi)$.
\\(3) \textbf{Shifting}: $x[m-m_0,n-n_0]\Leftrightarrow \exp(-i m_0 \omega_1) \exp(-j n_0 \omega_2) X(\omega_1,\omega_2)$.
\\(4) \textbf{Modulation}:  $\exp(i m \omega_{10}) \exp(j n \omega_{20}) x[m,n]\Leftrightarrow  X(\omega_1-\omega_{10},\omega_2-\omega_{20})$.
\\(5) \textbf{Energy Conservation}: $\sum_{-\infty}^{\infty} \sum_{-\infty}^{\infty} |x[m,n]|^2= \frac{1}{2\pi}\frac{1}{2\pi}\int_{-\pi}^{\pi}\int_{-\pi}^{\pi} |X(\omega_1,\omega_2)|^2 \ud \omega_1 \ud \omega_2$, where $|X(\omega_1,\omega_2)|^2= X(\omega_1,\omega_2) \bar{X}^*(\omega_1,\omega_2)$.

Now, it is strait forward to generalize the single orthant 2D-DTFT case and obtain the single orthant MD-DTFT.
Let $x[n_1,\cdots,n_M]$ be a non-periodic and real function of time, then the ND discrete time Fourier transform (ND-DTFT) of $x[n_1,\cdots,n_M]$, we define as
\begin{multline}
X(\omega_1,\cdots,\omega_M)=\sum_{n_1=-\infty}^{\infty}\cdots \sum_{n_M=-\infty}^{\infty} x[n_1,\cdots,n_M] \\ \times \exp(-j_1\omega_1 n_1-\cdots-j_M\omega_M n_M)  \label{Nddft_pro1}
\end{multline}
such that
\begin{equation}
 j_k.j_l=-1, \label{Nddft_pro3}
\end{equation}
where $j_1$ to $j_M$ are purely imaginary numbers and represent phase corresponding to frequencies $\omega_1$ to $\omega_M$, respectively. Since, there are $M$ frequencies $\omega_1$ to $\omega_M$ and integration of frequency is phase, hence it is logical and natural to define $M$ phases corresponding to $M$ frequencies. Here, we define $M$ conjugations for each $j_1$ to $j_M$, respectively.


It is to be noted that all the above discussions have been done with MD-FS, MD-DTFT and associated MD-analytic signal. It is very easy to extend this discussion for all the variants of Hilbert transforms and Fourier theory including Fourier transform (FT), DTFS, DFT and FFT.
\section{Numerical results and discussions}
To demonstrate that the HT and associated AS, defined through the FT in this paper, satisfy all the properties, we performed the following simulations and compare the results with some Hilbert transforms available in literature.
\subsection{Amplitude modulation, the Hilbert transform and Fourier representation} 
\begin{figure}[!t]
\centering
\includegraphics[angle=0,width=0.5\textwidth]{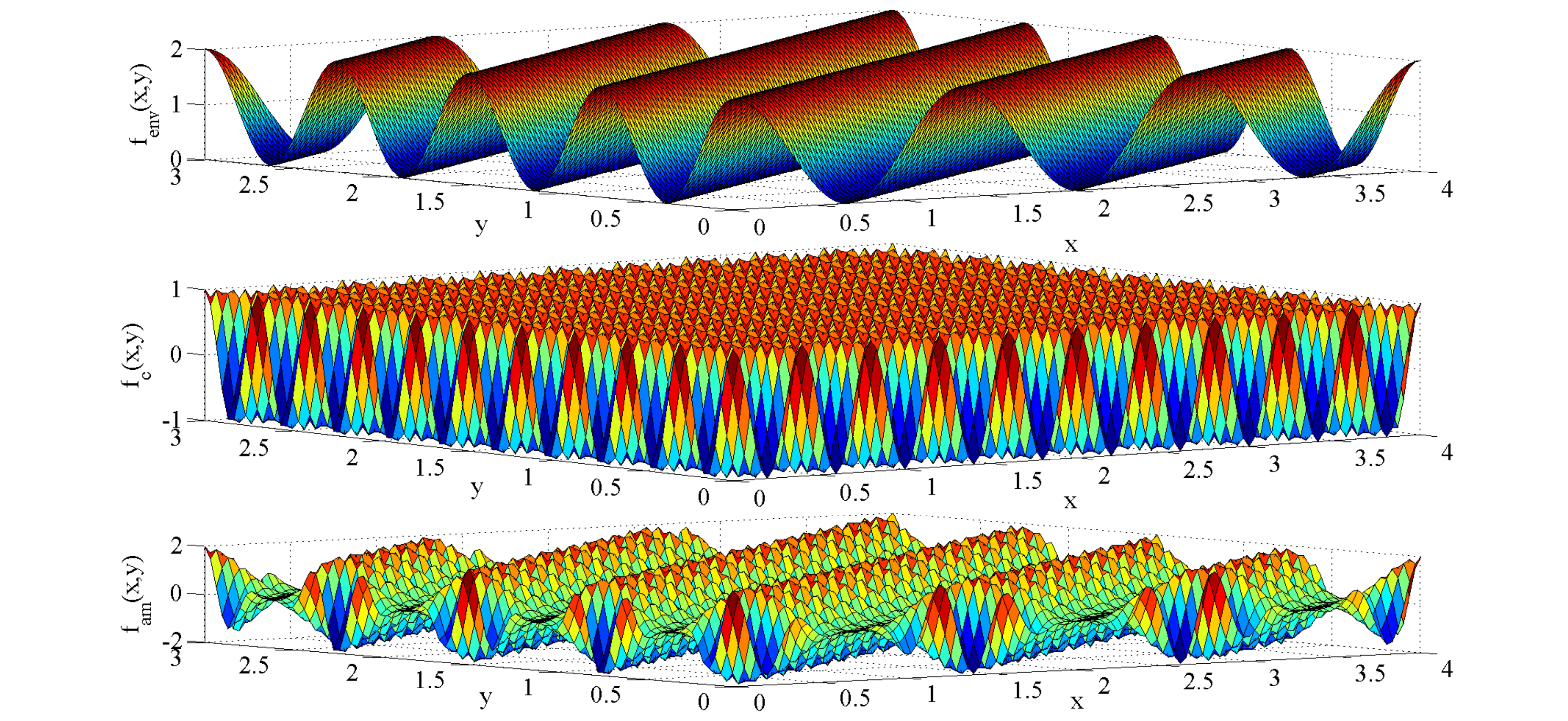}
\captionof{figure}{The envelope signal $f_{env}(x,y)=[1+f_m(x,y)]$ (top), carrier signal $f_c(x,y)$ (middle) and AM signal $f_{am}(x,y)=f_{env}(x,y)f_c(x,y)$ (bottom).}
\label{fig:AMFigure}
\end{figure}

\begin{figure}[!t]
\centering
\includegraphics[angle=0,width=0.5\textwidth]{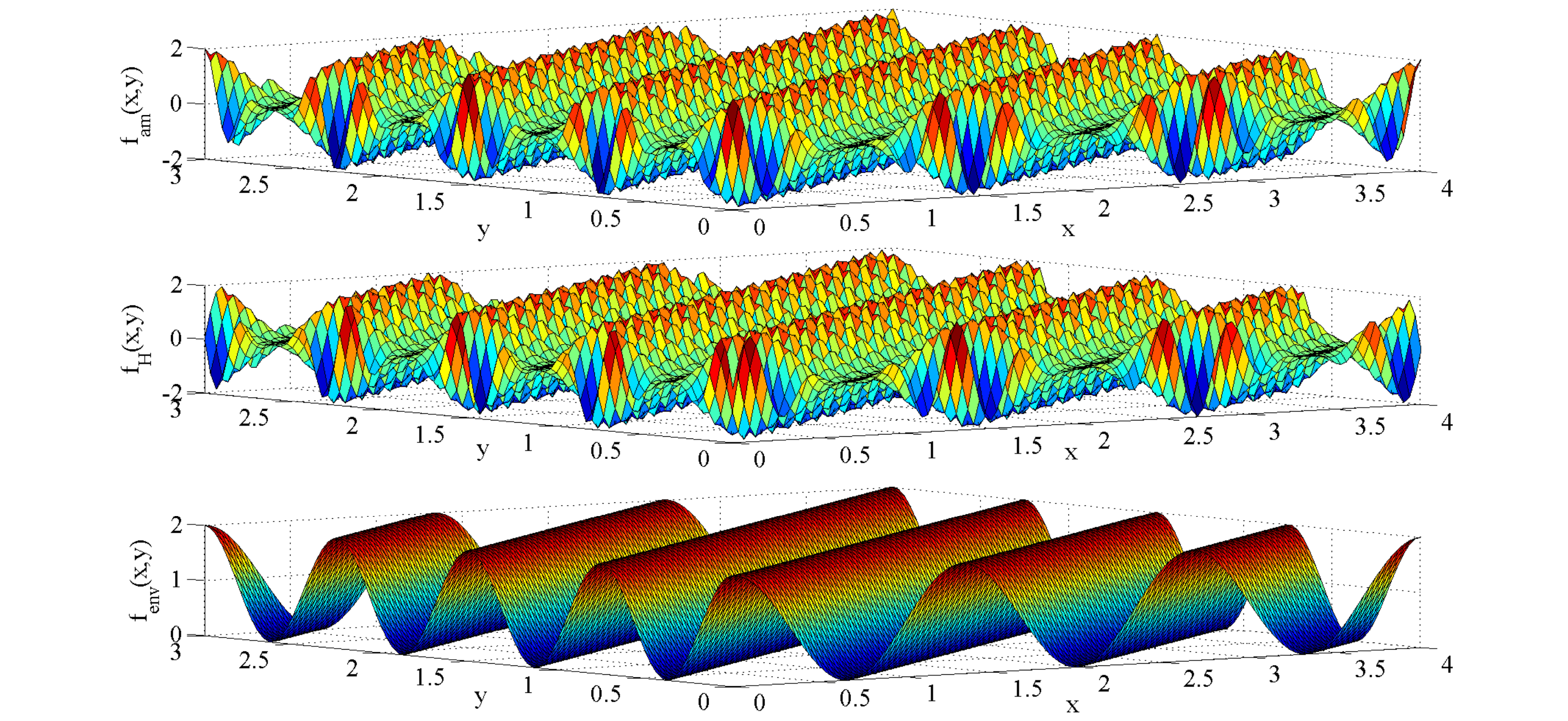}
\captionof{figure}{The recovered AM signal, $f_{am}(x,y)$, from the FS (top), the HT of recovered AM signal $f_{H}(x,y)$ (middle) and perfectly recovered envelope signal $f_{env}(x,y)$ (bottom).}
\label{fig:AMrFigure}
\end{figure}

Let $f_m(x,y)=A_m\cos(3\omega_1 x-4\omega_2 y)$ be a modulating signal and $f_c(x,y)=A_c\cos(10\omega_1 x+8\omega_2 y)$ be a carrier signal, Figure \ref{fig:AMFigure} (middle), with $\omega_1=\frac{2\pi}{T_x}$, $\omega_2=\frac{2\pi}{T_y}$, $T_x=3$, $T_y=4$, $A_c=1$, $A_m=1$ and spatial sampling frequency $\omega_s=2\pi f_s=2\pi\times30$ (rad/m). The AM modulated signal can be written as $f_{AM}(x,y)=[A_m+f_m(x,y)]A_c\cos(10\omega_1 x+8\omega_2 y)$, Figure \ref{fig:AMFigure} (bottom). The envelope of this modulated signal is given by $f_{env}(x,y)=[A_m+f_m(x,y)]$, which is shown in Figure \ref{fig:AMFigure} (top).
Figure \ref{fig:AMrFigure} shows the perfectly recovered AM signal (top), the HT of AM signal (middle) and envelope signal (bottom).
\subsection{Examples of 2D Analytic Signals}
 Consider the two-dimensional harmonic signal $f(x_1,x_2)=\cos(\omega_1 x_1)\cos(\omega_2 x_2)$. Using \eqref{FS2D_1}, the Fourier series expansion of this signal is given by $f(x_1,x_2)=\frac{1}{2}\cos(\omega_1 x_1+\omega_2 x_2)+\frac{1}{2}\cos(\omega_1 x_1-\omega_2 x_2)$, which implies that $f(x_1,x_2)$ signal is sum of two sinusoidal signals of phases $(\omega_1 x_1+\omega_2 x_2)$ and $(\omega_1 x_1-\omega_2 x_2)$, respectively. Using \eqref{1ddft4_1} or \eqref{FS2D_5_1}, and \eqref{FS2D_6} we obtain 2D-AS $z(x_1,x_2)=\frac{1}{2}e^{j(\omega_1 x_1+\omega_2 x_2)}+\frac{1}{2}e^{j(\omega_1 x_1-\omega_2 x_2)}$. This 2D-AS $z(x_1,x_2)$ has amplitude dependent on $x_1, x_2$ and a nonlinear phase. This shows that the superposition of constant amplitude and linear phase signals result in a signal with variable amplitude and a nonlinear phase. Whereas, the 2D-AS with single-quadrant spectra~\cite{NDFHT8} is given by $\psi(x_1,x_2)=e^{j(\omega_1 x_1+\omega_2 x_2)}$ that shows $f(x_1,x_2)$ has a constant amplitude and a linear phase in both coordinates.

 Other examples are presented in Table~\ref{table:HT_table} and Table~\ref{table:AS_table} for the comparative study of the proposed Hilbert transforms (HT) $\hat{f}(x_1,x_2)$, which is based on the Fourier theory and phase delay, with the four Hilbert transforms~\cite{NDFHT8}, namely the partial HT (PHT) in $x_1$ direction $\hat{f}_{x_1}(x_1,x_2)$ with 2D-AS $z(x_1,x_2)=f(x_1,x_2)+j\hat{f}_{x_1}(x_1,x_2)$, PTH in $x_2$ direction $\hat{f}_{x_2}(x_1,x_2)$ with 2D-AS $z(x_1,x_2)=f(x_1,x_2)+j\hat{f}_{x_2}(x_1,x_2)$, total HT (THT) $\hat{f}_{T}(x_1,x_2)$ with 2D-AS $z(x_1,x_2)=f(x_1,x_2)+j\hat{f}_{T}(x_1,x_2)$, and single-orthant HT (SOHT) $\hat{f}_{SO}(x_1,x_2)$ with 2D-AS $z(x_1,x_2)=f(x_1,x_2)-\hat{f}_{T}(x_1,x_2)+j[\hat{f}_{x_1}(x_1,x_2)+\hat{f}_{x_2}(x_1,x_2)]$. From these examples it is clear that the only proposed HT with 2D-AS, presented in Table~\ref{table:AS_table}, is proving correct amplitude and phase in all the cases.

\begin{figure*}[!t]
\centering
\includegraphics[angle=0,width=1\textwidth]{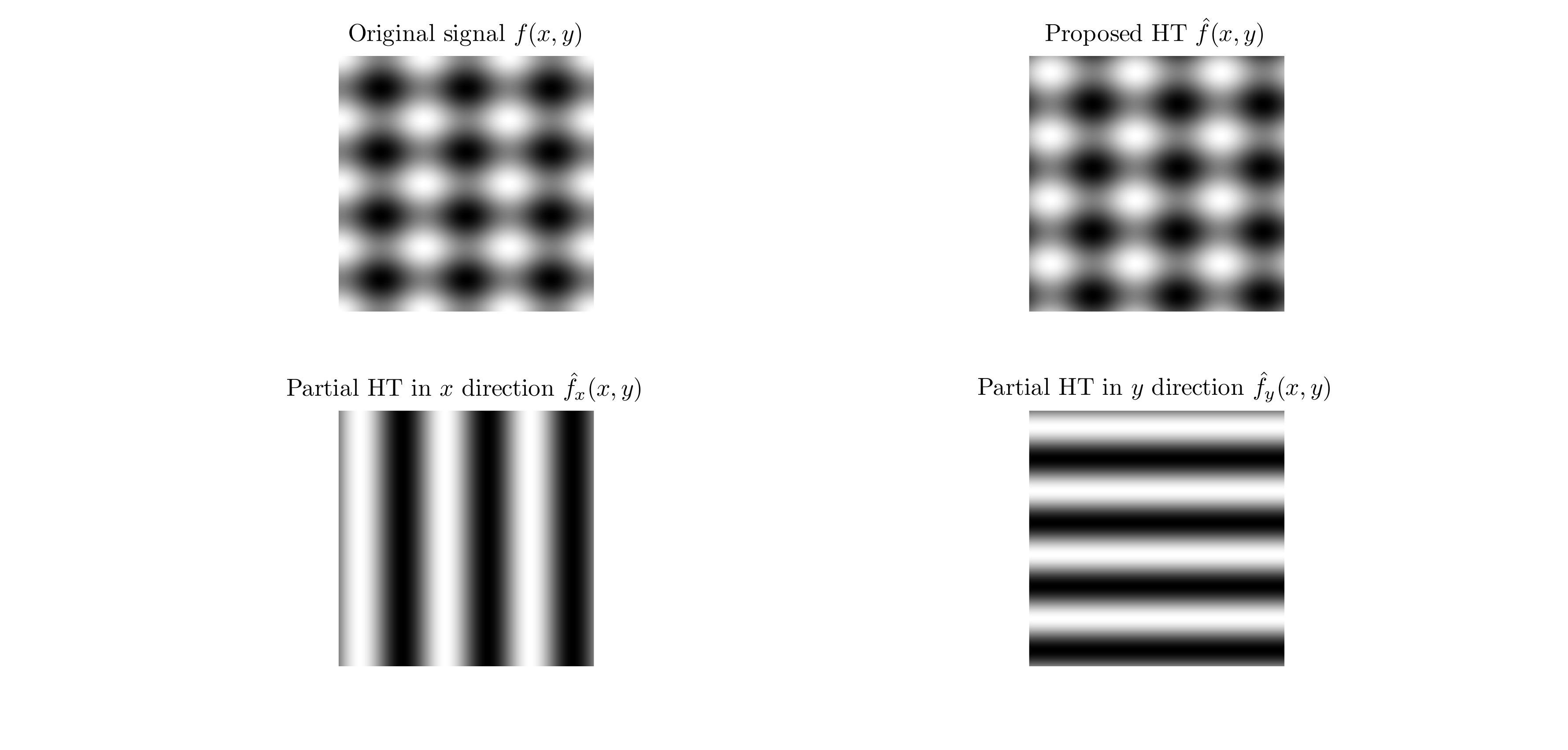}
\captionof{figure}{Original signal $f(x,y)$ and its proposed HT $\hat{f}(x,y)$, partial HT in $x$ direction $\hat{f}_x(x,y)$ and partial HT in $y$ direction $\hat{f}_y(x,y)$.}
\label{fig:ProposedHT_PHT}
\end{figure*}

 Figure \ref{fig:ProposedHT_PHT} shows a signal $f(x,y)$ and its proposed HT $\hat{f}(x,y)$, partial HT in $x$ direction $\hat{f}_x(x,y)$ and partial HT in $y$ direction $\hat{f}_y(x,y)$. This example clearly demonstrate that the proposed HT is not oriented to any direction, unlike PHT.

From the Table~\ref{table:HT_table} and Table~\ref{table:AS_table}, we observe that the proposed HT $\hat{f}(x,y)$ is closest to PTH in $x$ direction $\hat{f}_x(x,y)$. To observe the difference between these two, we rewrite \eqref{FS2D_1} as
\begin{multline}
f(x,y)= a_{0,0}+\sum_{l=1}^{\infty} \sum_{k=1}^{\infty}\Big[ a_{k,l} \cos(k\omega_1x+l\omega_2y) \\+ b_{k,l} \sin(k\omega_1x+l\omega_2y) \Big]  + \\+ \sum_{k=1}^{\infty} \Big[ a_{k,0} \cos(k\omega_1x) + b_{k,0} \sin(k\omega_1x) \Big]
\\+ \sum_{l=1}^{\infty} \Big[ a_{0,l} \cos(l\omega_2y) + b_{0,l} \sin(l\omega_2y) \Big]\\
+\sum_{l=-\infty}^{-1} \sum_{k=1}^{\infty}\Big[ a_{k,l} \cos(k\omega_1x+l\omega_2y) \\+ b_{k,l} \sin(k\omega_1x+l\omega_2y) \Big], \label{examplFS2D_1}
\end{multline}
and obtain a partial Hilbert transform in $x$ direction as
 \begin{multline}
\hat{f}_x(x,y)= \sum_{l=1}^{\infty} \sum_{k=1}^{\infty}\Big[ a_{k,l} \sin(k\omega_1x+l\omega_2y) \\- b_{k,l} \cos(k\omega_1x+l\omega_2y) \Big]  + \\+ \sum_{k=1}^{\infty} \Big[ a_{k,0} \sin(k\omega_1x) - b_{k,0} \cos(k\omega_1x) \Big]\\
+\sum_{l=-\infty}^{-1} \sum_{k=1}^{\infty}\Big[ a_{k,l} \sin(k\omega_1x+l\omega_2y) \\- b_{k,l} \cos(k\omega_1x+l\omega_2y) \Big]. \label{examplFS2D_2}
\end{multline}
Equations \eqref{FS2D_5_1} and \eqref{examplFS2D_2} provide clear differences between the proposed HT and PHT in $x$ direction.
\begin{table*}[!t]
\caption{The partial Hilbert transform (PHT) in $x_1$ direction $\hat{f}_{x_1}(x_1,x_2)$, PTH in $x_2$ direction $\hat{f}_{x_2}(x_1,x_2)$, total HT (THT) $\hat{f}_{T}(x_1,x_2)$ and single-orthant HT (SOHT) $\hat{f}_{SO}(x_1,x_2)$.} 
\centering 
\begin{tabular}{|c|c|c|c|c|}
  \hline
  function ${f}(x_1,x_2)$ & $\hat{f}_{x_1}(x_1,x_2)$ & $\hat{f}_{x_2}(x_1,x_2)$ & $\hat{f}_{T}(x_1,x_2)$ & $\hat{f}_{SO}(x_1,x_2)$ \\ \hline
  constant ($a_0$) & 0 & 0 & 0 & 0 \\\hline
  $\sin(\omega_2 x_2)\sin(\omega_1 x_1)$ & $-\sin(\omega_2 x_2)\cos(\omega_1 x_1)$ & $-\cos(\omega_2 x_2)\sin(\omega_1 x_1)$ & $\cos(\omega_2 x_2)\cos(\omega_1 x_1)$ & $-\sin(\omega_1 x_1+\omega_2 x_2)$ \\ \hline

  $\cos(\omega_2 x_2)\sin(\omega_1 x_1)$ & $-\cos(\omega_2 x_2)\cos(\omega_1 x_1)$ & $\sin(\omega_2 x_2)\sin(\omega_1 x_1)$ & $-\sin(\omega_2 x_2)\cos(\omega_1 x_1)$ & $-\cos(\omega_1 x_1+\omega_2 x_2)$  \\ \hline

  $\sin(\omega_2 x_2)\cos(\omega_1 x_1)$ & $\sin(\omega_2 x_2)\sin(\omega_1 x_1)$ & $-\cos(\omega_2 x_2)\cos(\omega_1 x_1)$ & $-\cos(\omega_2 x_2)\sin(\omega_1 x_1)$ & $-\cos(\omega_1 x_1+\omega_2 x_2)$  \\ \hline

  $\cos(\omega_1 x_1+\omega_2 x_2)$ & $\sin(\omega_1 x_1+\omega_2 x_2)$ & $\sin(\omega_1 x_1+\omega_2 x_2)$ & $-\cos(\omega_1 x_1+\omega_2 x_2)$ & $2\sin(\omega_1 x_1+\omega_2 x_2)$  \\\hline

  $\cos(\omega_1 x_1-\omega_2 x_2)$ & $\sin(\omega_1 x_1-\omega_2 x_2)$ & $-\sin(\omega_1 x_1-\omega_2 x_2)$ & $\cos(\omega_1 x_1-\omega_2 x_2)$ & 0 \\\hline

  $\sin(\omega_1 x_1+\omega_2 x_2)$ & $-\cos(\omega_1 x_1+\omega_2 x_2)$ & $-\cos(\omega_1 x_1+\omega_2 x_2)$ & $-\sin(\omega_1 x_1+\omega_2 x_2)$ & $-2\cos(\omega_1 x_1+\omega_2 x_2)$ \\\hline

  $\sin(\omega_1 x_1-\omega_2 x_2)$ & $-\cos(\omega_1 x_1-\omega_2 x_2)$ & $\cos(\omega_1 x_1-\omega_2 x_2)$ & $\sin(\omega_1 x_1-\omega_2 x_2)$ & 0 \\\hline

  $\cos(\omega_1 x_1)$ & $\sin(\omega_1 x_1)$ & 0 & 0 & $\sin(\omega_1 x_1)$ \\\hline
  $\cos(\omega_2 x_2)$ & 0 & $\sin(\omega_2 x_2)$ & 0 & $\sin(\omega_2 x_2)$ \\\hline
  $\sin(\omega_1 x_1)$ & $-\cos(\omega_1 x_1)$ & 0 & 0 & $-\cos(\omega_1 x_1)$ \\\hline
  $\sin(\omega_2 x_2)$ & 0 & $-\cos(\omega_2 x_2)$ & 0 & $-\cos(\omega_2 x_2)$ \\
  \hline
\end{tabular}
\label{table:HT_table} 
\end{table*}
\begin{table*}[!t]
\caption{The proposed Fourier theory and phase delay based 2D HT $\hat{f}(x_1,x_2)$ and analytic signal (2D-AS) $z(x_1,x_2)$.} 
\centering 
\begin{tabular}{|c|c|c|}
  \hline
  function ${f}(x_1,x_2)$ & proposed $\hat{f}(x_1,x_2)$ & proposed 2D-AS $z(x_1,x_2)$\\ \hline
  constant ($a_0$) & 0 & $a_0$ \\
  $\sin(\omega_2 x_2)\sin(\omega_1 x_1)$ & $-\sin(\omega_2 x_2)\cos(\omega_1 x_1)$ & $\frac{1}{2}[e^{j(\omega_1 x_1-\omega_2 x_2)}-e^{j(\omega_1 x_1+\omega_2 x_2)}]$\\
  $\cos(\omega_2 x_2)\sin(\omega_1 x_1)$ & $-\cos(\omega_2 x_2)\cos(\omega_1 x_1)$ & $\frac{1}{2}[e^{j(\omega_1 x_1-\omega_2 x_2-\frac{\pi}{2})}+e^{j(\omega_1 x_1+\omega_2 x_2-\frac{\pi}{2})}]$\\
  $\sin(\omega_2 x_2)\cos(\omega_1 x_1)$ & $\sin(\omega_2 x_2)\sin(\omega_1 x_1)$ & $\frac{1}{2}[e^{j(\omega_1 x_1-\omega_2 x_2+\frac{\pi}{2})}+e^{j(\omega_1 x_1+\omega_2 x_2-\frac{\pi}{2})}]$\\
  $\cos(\omega_1 x_1+\omega_2 x_2)$ & $\sin(\omega_1 x_1+\omega_2 x_2)$ & $e^{j(\omega_1 x_1+\omega_2 x_2)}$\\
  $\cos(\omega_1 x_1-\omega_2 x_2)$ & $\sin(\omega_1 x_1-\omega_2 x_2)$ & $e^{j(\omega_1 x_1-\omega_2 x_2)}$ \\
  $\sin(\omega_1 x_1+\omega_2 x_2)$ & $-\cos(\omega_1 x_1+\omega_2 x_2)$ & $e^{j(\omega_1 x_1+\omega_2 x_2-\frac{\pi}{2})}$ \\
  $\sin(\omega_1 x_1-\omega_2 x_2)$ & $-\cos(\omega_1 x_1-\omega_2 x_2)$ & $e^{j(\omega_1 x_1-\omega_2 x_2-\frac{\pi}{2})}$ \\
  $\cos(\omega_1 x_1)$ & $\sin(\omega_1 x_1)$ & $e^{j(\omega_1 x_1)}$  \\
  $\cos(\omega_2 x_2)$ & $\sin(\omega_2 x_2)$ & $e^{j(\omega_2 x_2)}$ \\
  $\sin(\omega_1 x_1)$ & $-\cos(\omega_1 x_1)$ & $e^{j(\omega_1 x_1-\frac{\pi}{2})}$ \\
  $\sin(\omega_2 x_2)$ & $-\cos(\omega_2 x_2)$ & $e^{j(\omega_2 x_2-\frac{\pi}{2})}$ \\
  \hline
\end{tabular}
\label{table:AS_table} 
\end{table*}
\subsection{Edge detection}
\begin{figure*}[!t]
\centering
\includegraphics[angle=0,width=1\textwidth]{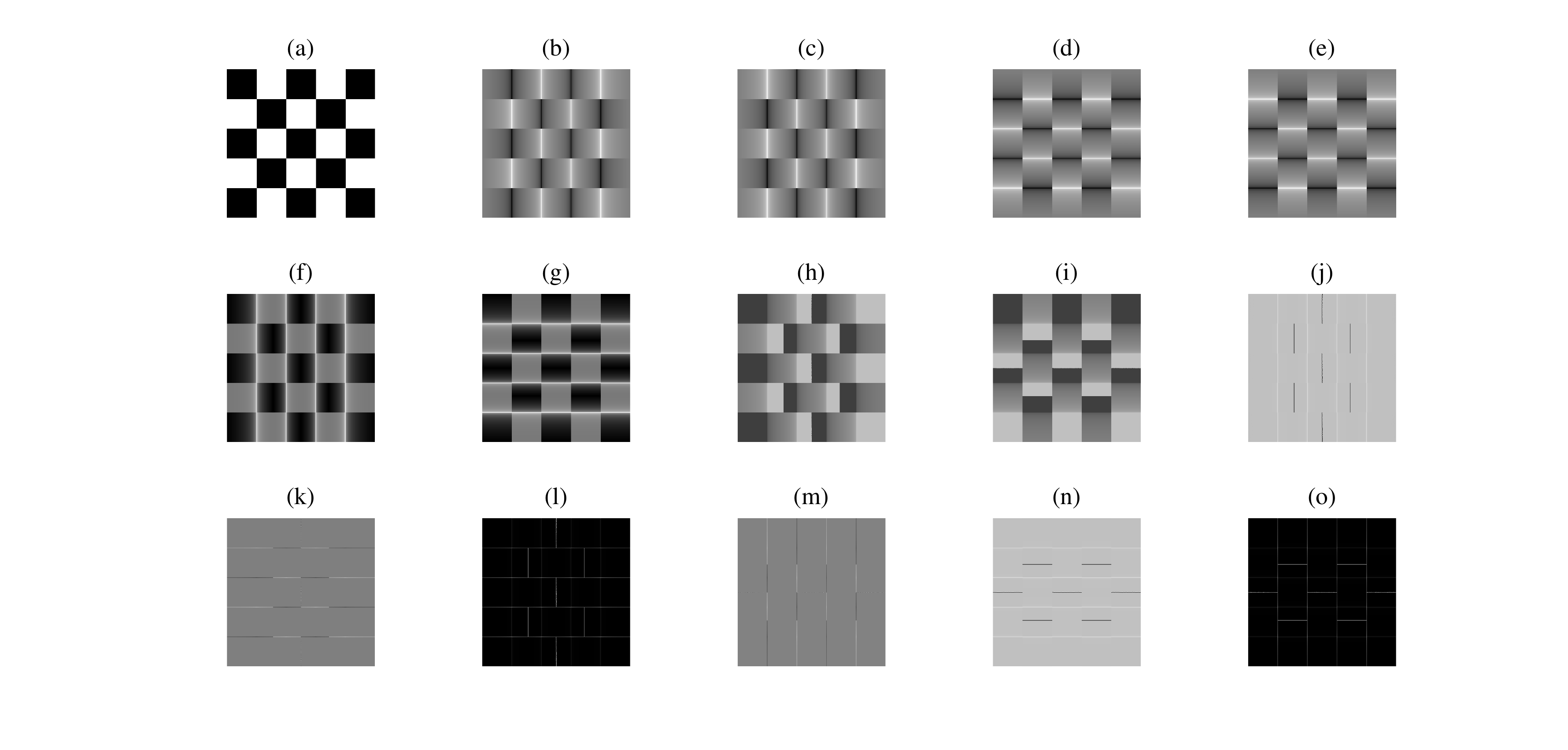}
\captionof{figure}{Directional HT (DHT) (a) Checkerboard image ${x}[m,n]$ (b) DHT $\hat{x}_{14}[m,n]$ (c) DHT $\hat{x}_{23}[m,n]$ (d) DHT $\hat{x}_{12}[m,n]$ (e) DHT $\hat{x}_{34}[m,n]$ (f) analytic amplitude $|{z}_{14}[m,n]|$ (g) analytic amplitude $|{z}_{12}[m,n]|$ (h) analytic phase $\angle {z}_{14}[m,n]$ (AP-1-4) (i) analytic phase $\angle {z}_{12}[m,n]$ (AP-1-2) (j) gradient $\phi_m[m,n]$ of AP-1-4 (k) gradient $\phi_n[m,n]$ of AP-1-4 (l) $\sqrt{\phi^2_m[m,n]+\phi^2_n[m,n]}$ of quadrant 1-4 (m) gradient $\phi_m[m,n]$ of AP-1-2 (n) gradient $\phi_n[m,n]$ of AP-1-2 (o) $\sqrt{\phi^2_m[m,n]+\phi^2_n[m,n]}$ of quadrant 1-2.}
\label{fig:Checkerboard}
\end{figure*}

\begin{figure*}[!t]
\centering
\includegraphics[angle=0,width=1\textwidth]{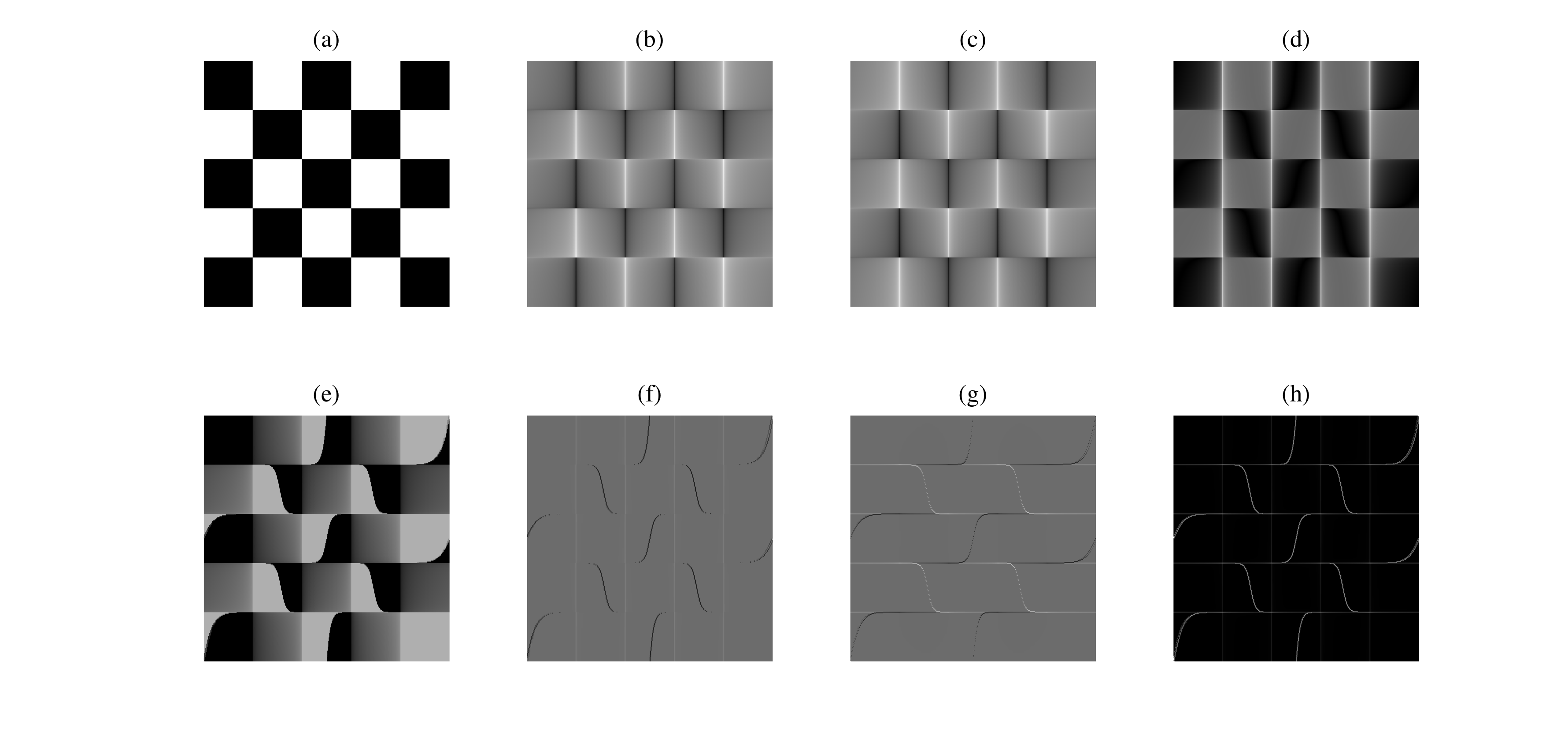}
\captionof{figure}{Proposed HT (a) Checkerboard image ${x}[m,n]$ (b) HT $\hat{x}_{14}[m,n]$ (c) HT $\hat{x}_{23}[m,n]$ (d) analytic amplitude $|{z}_{14}[m,n]|$ (e) analytic phase $\angle {z}_{14}[m,n]$ (AP-1-4) (f) gradient $\phi_m[m,n]$ of AP-1-4 (g) gradient $\phi_n[m,n]$ of AP-1-4 (h) $\sqrt{\phi^2_m[m,n]+\phi^2_n[m,n]}$ of quadrant 1-4.}
\label{fig:checkerboardProposedHT}
\end{figure*}

\begin{figure*}[!t]
\centering
\includegraphics[angle=0,width=1\textwidth]{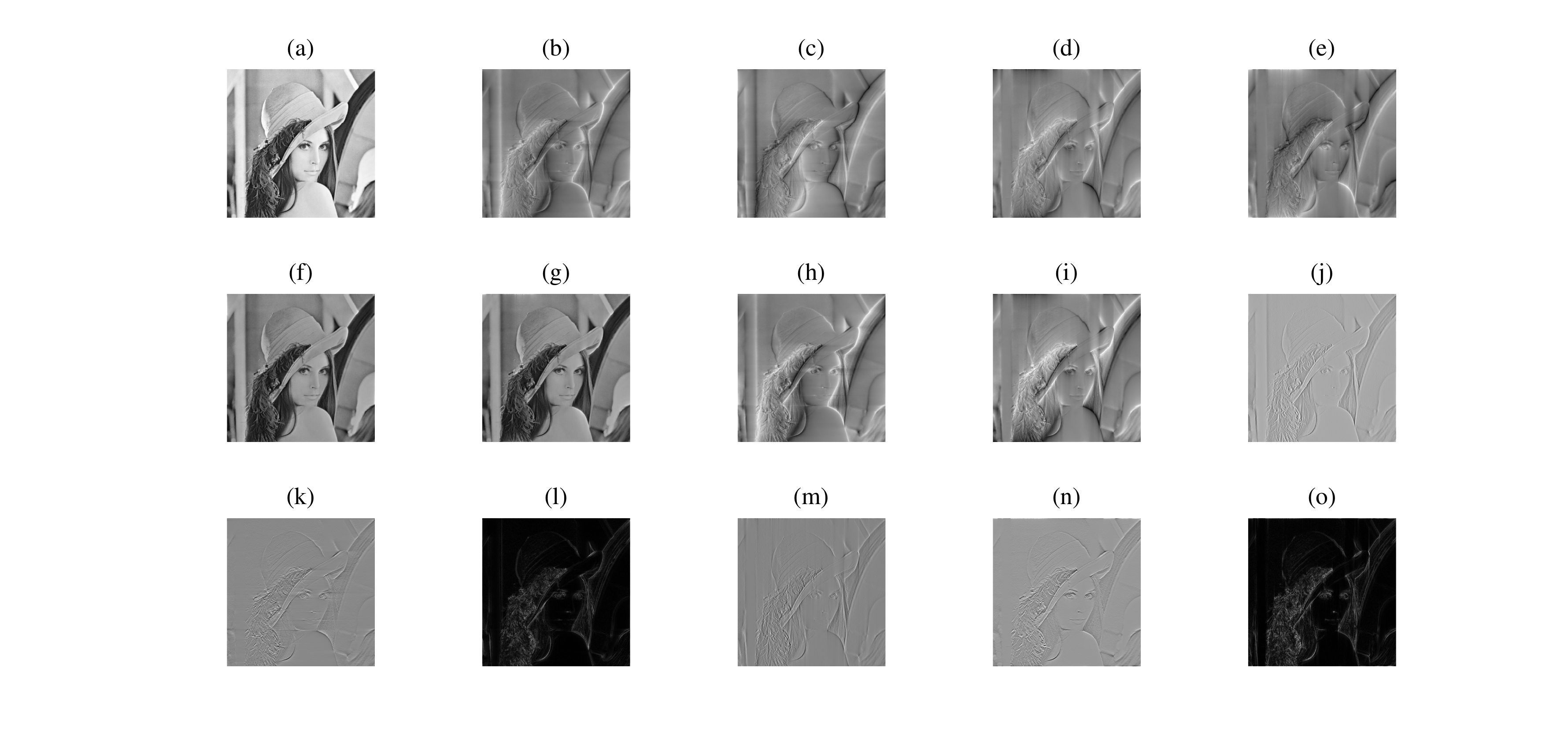}
\captionof{figure}{Directional HT (DHT) (a) Lena image ${x}[m,n]$ (b) DHT $\hat{x}_{14}[m,n]$ (c) DHT $\hat{x}_{23}[m,n]$ (d) DHT $\hat{x}_{12}[m,n]$ (e) DHT $\hat{x}_{34}[m,n]$ (f) analytic amplitude $|{z}_{14}[m,n]|$ (g) analytic amplitude $|{z}_{12}[m,n]|$ (h) analytic phase $\angle {z}_{14}[m,n]$ (AP-1-4) (i) analytic phase $\angle {z}_{12}[m,n]$ (AP-1-2) (j) gradient $\phi_m[m,n]$ of AP-1-4 (k) gradient $\phi_n[m,n]$ of AP-1-4 (l) $\sqrt{\phi^2_m[m,n]+\phi^2_n[m,n]}$ of quadrant 1-4 (m) gradient $\phi_m[m,n]$ of AP-1-2 (n) gradient $\phi_n[m,n]$ of AP-1-2 (o) $\sqrt{\phi^2_m[m,n]+\phi^2_n[m,n]}$ of quadrant 1-2.}
\label{fig:Lena}
\end{figure*}

\begin{figure*}[!t]
\centering
\includegraphics[angle=0,width=1\textwidth]{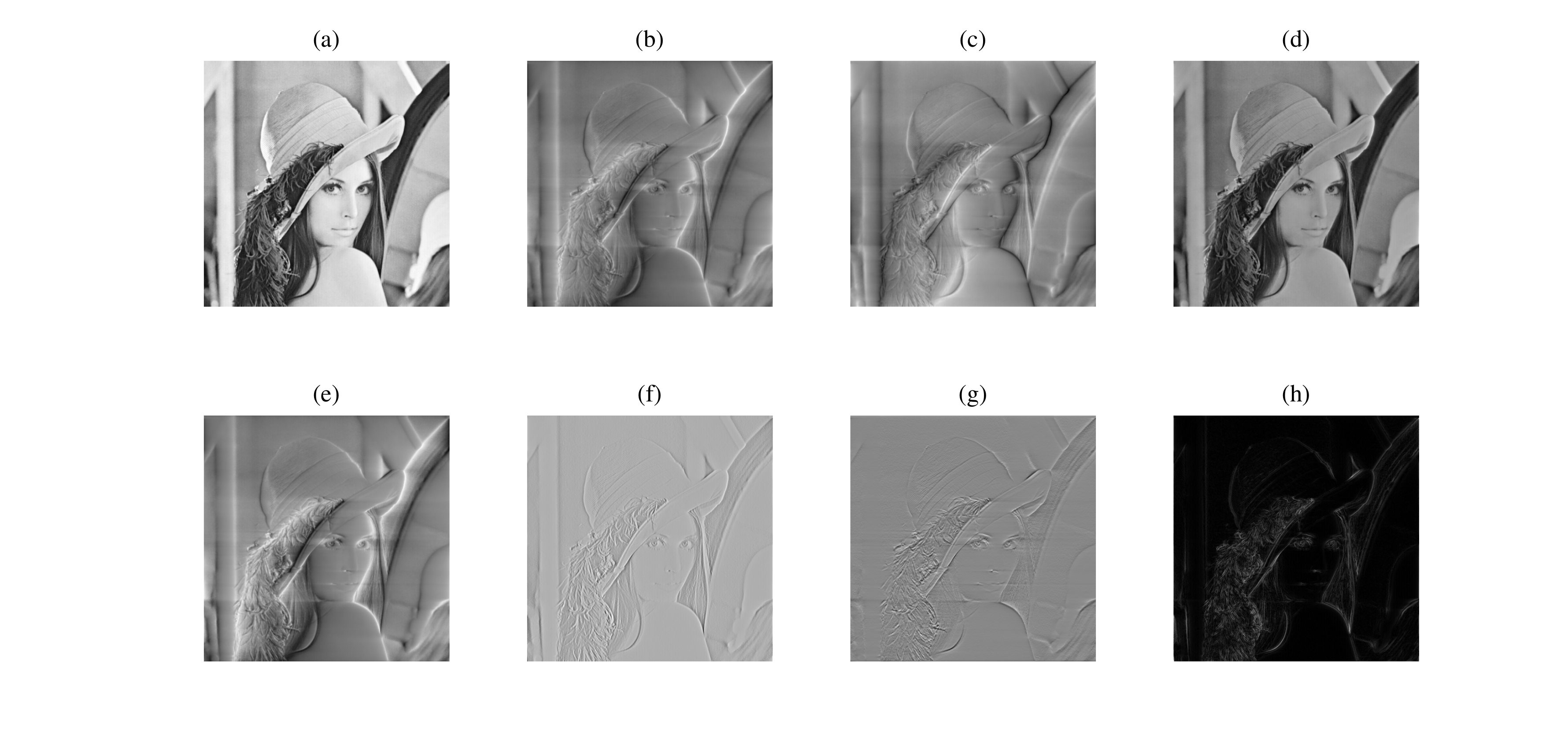}
\captionof{figure}{Proposed HT (a) Checkerboard image ${x}[m,n]$ (b) HT $\hat{x}_{14}[m,n]$ (c) HT $\hat{x}_{23}[m,n]$ (d) analytic amplitude $|{z}_{14}[m,n]|$ (e) analytic phase $\angle {z}_{14}[m,n]$ (AP-1-4) (f) gradient $\phi_m[m,n]$ of AP-1-4 (g) gradient $\phi_n[m,n]$ of AP-1-4 (h) $\sqrt{\phi^2_m[m,n]+\phi^2_n[m,n]}$ of quadrant 1-4.}
\label{fig:lena512ProposedHT}
\end{figure*}

The Hilbert transform can be seen as an edge detectors. The Figure \ref{fig:Checkerboard} and \ref{fig:Lena} show examples of gray level checkerboard image and natural image, respectively. We have calculated the directional HT (DHT), corresponding to first and fourth quadrant (DHT-1-4), first and second quadrant (DHT-1-2), using FFT based algorithm. It appears that these Hilbert transforms act as edge detection steered in the x and y directions. If we compare the two Hilbert components, we can see different kind of edges responding to the x and y directions. To observe the differences between the directional HT and proposed HT, we evaluated the proposed HT, AS, analytic phase and gradient of same gray level checkerboard image and natural image in Figure \ref{fig:checkerboardProposedHT} and \ref{fig:lena512ProposedHT}, respectively.
\subsection{2D-FDM}
\begin{figure*}[!t]
\centering
\includegraphics[angle=0,width=1\textwidth]{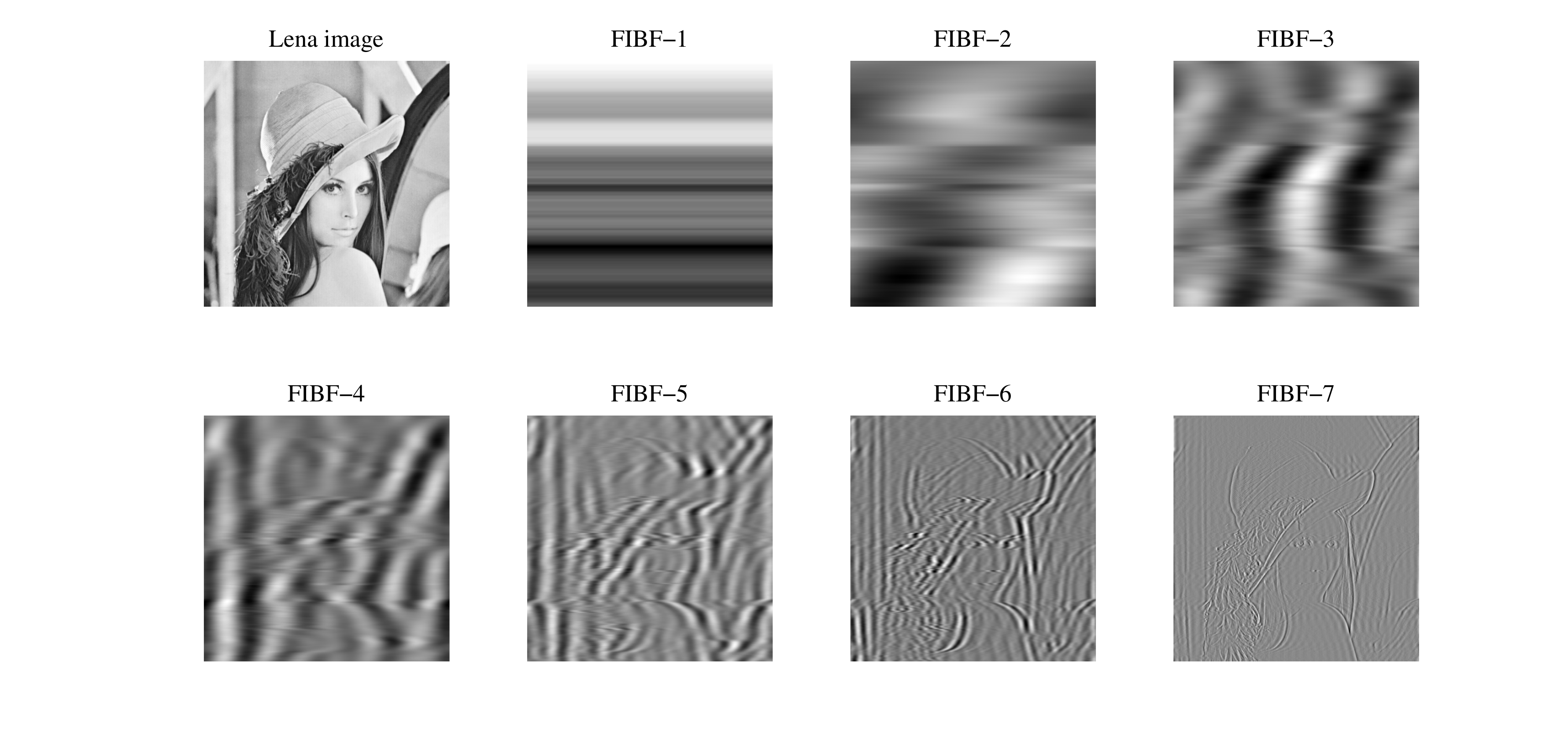}
\captionof{figure}{Lena image and its decomposition, w.r.t. $\omega_1$ by \eqref{FDM_eq261}, to orthogonal 2D-FIBFs 1-7 in order of increasing frequency components by 2D-FDM. FIBF-1 is lowest component and FIBF-7 is highest frequency component. Lena image is sum of all FIBFs 1-7.}
\label{fig:Lena_FDM14}
\end{figure*}

\begin{figure*}[!t]
\centering
\includegraphics[angle=0,width=1\textwidth]{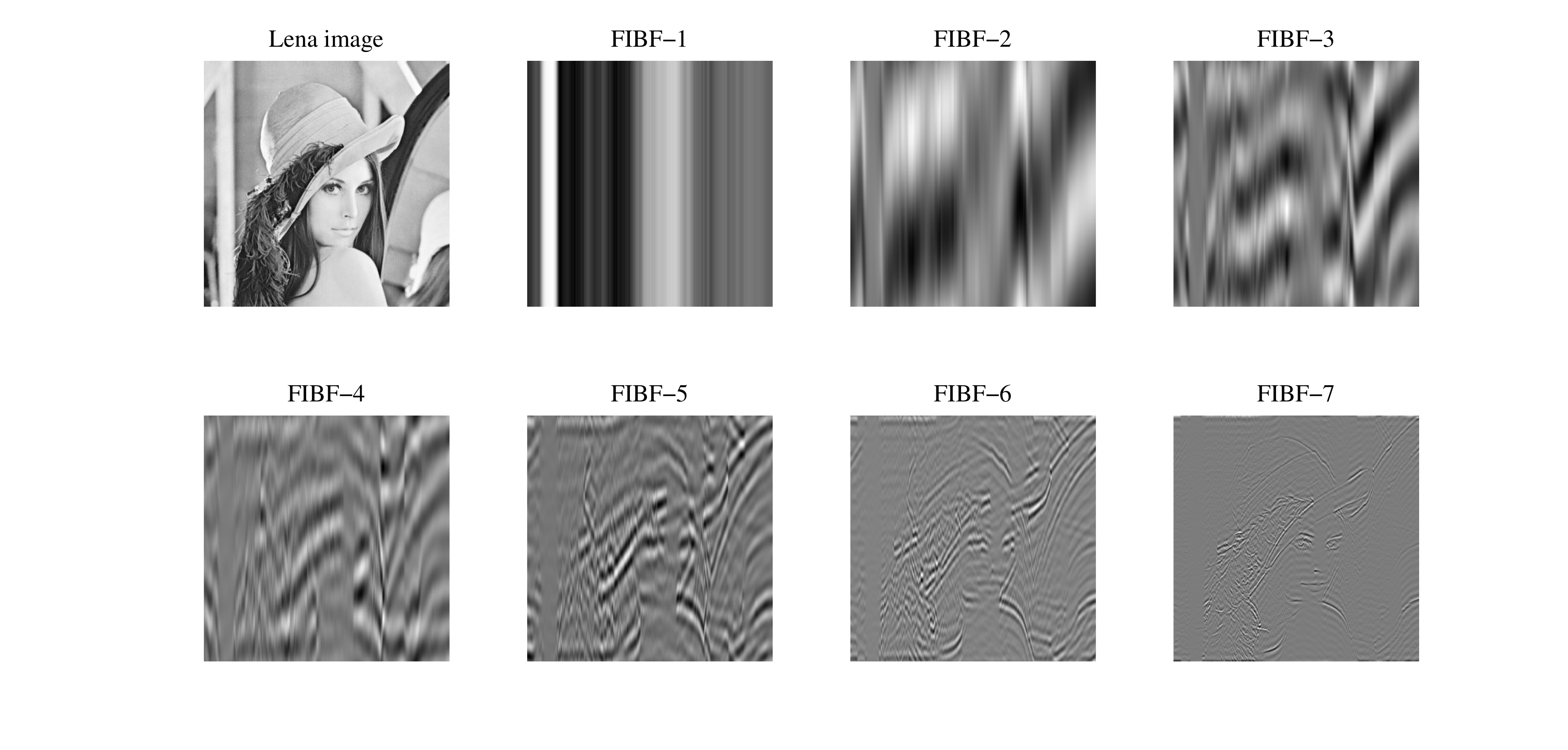}
\captionof{figure}{Lena image and its decomposition, w.r.t. $\omega_2$, to orthogonal 2D-FIBFs 1-7 in increasing frequency components by 2D-FDM. FIBF-1 is lowest frequency component and FIBF-7 is highest frequency component. Lena image is sum of all FIBFs 1-7.}
\label{fig:Lena_FDM}
\end{figure*}

\begin{figure*}[!t]
\centering
\includegraphics[angle=0,width=1\textwidth]{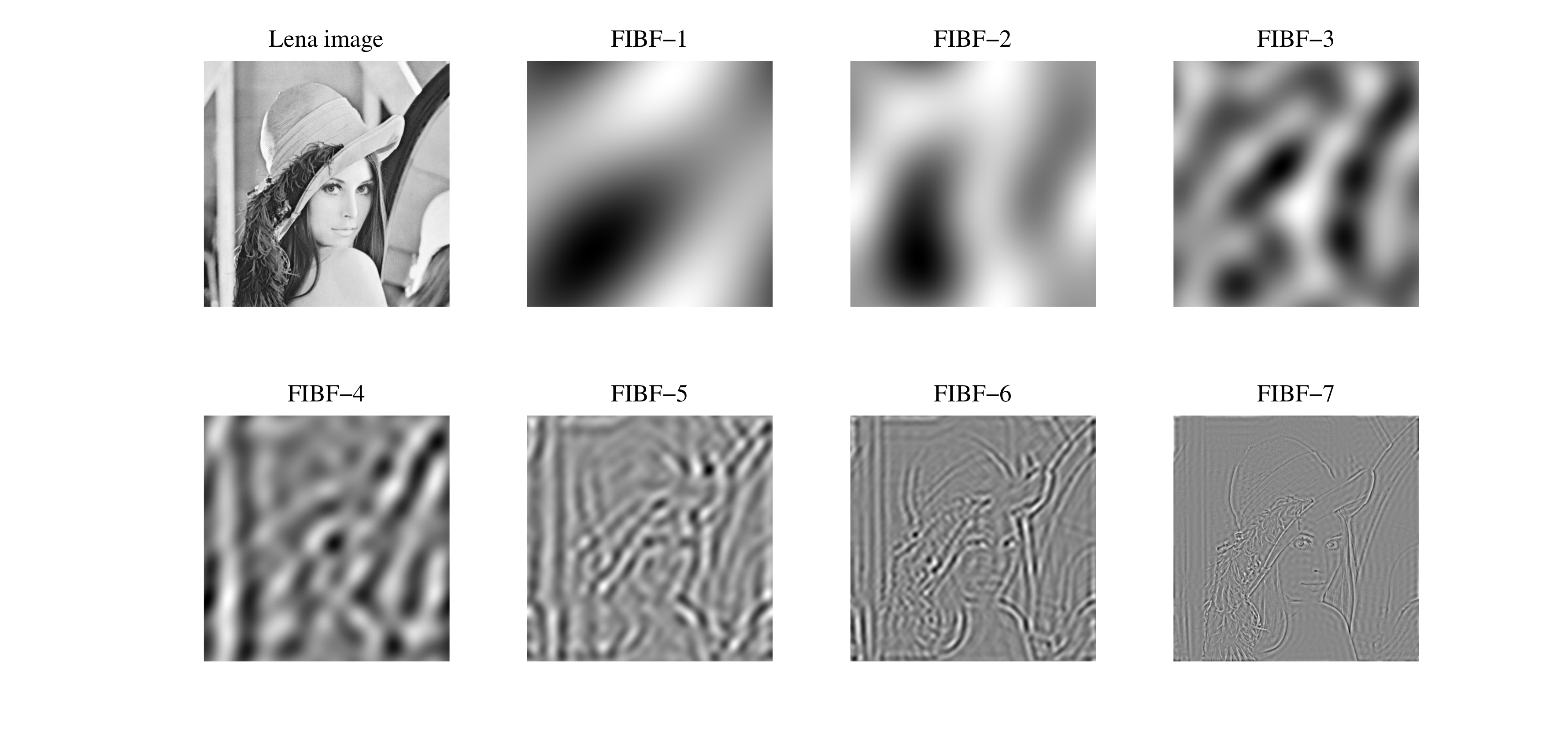}
\captionof{figure}{Lena image and its decomposition, using Fourier based zero-phase filtering (ZPF), to orthogonal 2D-FIBFs 1-7 in order of increasing frequency components by 2D-FDM. FIBF-1 is lowest frequency component and FIBF-7 is highest frequency component. Lena image is sum of all FIBFs 1-7.}
\label{fig:LenaZPF}
\end{figure*}

Lena image and its decomposition to orthogonal 2D FIBFs are shown in Figures \ref{fig:Lena_FDM14} and \ref{fig:Lena_FDM} by considering first and fourth quadrant (w.r.t. $\omega_1$), first and second quadrants (w.r.t. $\omega_2$), respectively. It appears that these decompositions are steered in the x and y directions. If we compare these two decompositions, we can see different kind of frequency components responding to the different, x and y, directions. Figure \ref{fig:LenaZPF} shows Lena image and its decomposition, using Fourier based zero-phase filtering, to orthogonal 2D-FIBFs 1-7 in order of increasing frequency components by 2D-FDM. FIBF-1 is lowest frequency component and FIBF-7 is highest frequency component. Lena image is sum of FIBF-1 to FIBF-7. From Figure \ref{fig:LenaZPF}, it is clear that this decomposition is not steered to any direction.
\subsection{Single orthant Fourier transform (SOFT) example}
To show how single orthant Fourier transform works, we take an example signal $x[m,n]$ which is shown in Table~\ref{table:Signal}.
The DFT of this signal is shown in Table~\ref{table:DFT}. The proposed single orthant DFT is shown in Table~\ref{table:SO_DFT} where first orthant DFT (SODFT) elements are shown in bold numbers and values in rest of the orthant are obtained by simple conjugations.
 The elements of DFT Table~\ref{table:DFT} can be obtained by the elements of SODFT Table~\ref{table:SO_DFT} from the relation $X[k,l]=(X_r[k,l]+X_{ij}[k,l])+j(X_i[k,l]+X_j[k,l])$.
\begin{table}[!t]
\caption{An example signal $x[m,n]$.} 
\centering 
\begin{tabular}{|c|c|c|}
  \hline
  1 & -7 & 30 \\ \hline
  40 & 5 & -6 \\ \hline
  -70 & 8 & -100 \\ \hline
\end{tabular}
\label{table:Signal} 
\end{table}
\begin{table}[!t]
\caption{The DFT $X[k,l]$ of an example signal $x[m,n]$ of Table~\ref{table:Signal}} 
\centering 
\begin{tabular}{|c|c|c|}
  \hline
  -99 & 6.00 - 71.01i & 6 + 71.01i \\ \hline
  85.50 - 174.07i & 54.00 + 27.71i & -91.50 - 139.43i \\\hline
  85.50 + 174.07i & -91.50 + 139.43i & 54.00 - 27.71i \\
  \hline
\end{tabular}
\label{table:DFT} 
\end{table}
\begin{table*}[!t]
\caption{The proposed single orthant DFT elements, of example signal $x[m,n]$ of Table~\ref{table:Signal}, in first orthant are shown in bold numbers and values in rest of the orthant are obtained by simple conjugations.} 
\centering 
\begin{tabular}{|c|c|c|}
  \hline
  $X_r,X_{ij},X_i,X_j$ & $X_r,X_{ij},X_i,X_j$ & $X_r,X_{ij},X_i,X_j$ \\ \hline
  \textbf{-99,  0,  0, 0} & \textbf{6.0000,         0,         0,  -71.0141} & 6.0000,         0,         0,   71.0141 \\ \hline
  \textbf{85.5000, 0, -174.0711, 0} & \textbf{-18.7500,   72.7500,  -55.8586,   83.5715} & -18.7500,  -72.7500,  -55.8586,  -83.5715 \\\hline
  85.5000,  0,  174.0711, 0 & -18.7500,  -72.7500,   55.8586,   83.5715 & -18.7500,   72.7500,   55.8586,  -83.5715 \\
  \hline
\end{tabular}
\label{table:SO_DFT} 
\end{table*}
\section{Conclusion}
In this study, we have proposed the Fourier frequency vector (FFV), inherently, associated with multidimensional Fourier transform. With the help of FFV, we have provided the physical meaning of so called negative frequencies in multidimensional Fourier transform (MDFT), which in turn provide multidimensional spatial and space-time series analysis. We have shown that the complex exponential representation of sinusoidal function always yields two frequencies, negative frequency corresponding to positive frequency and vice versa, in the multidimensional Fourier spectrum. Thus, using the MDFT, we have proposed multidimensional Hilbert transform (MDHT) and associated multidimensional analytic signal (MDAS) with following properties: (a) the extra and redundant positive, negative, or both frequencies, introduced due to complex exponential representation of multidimensional Fourier spectrum, are suppressed, (b) real part of MDAS is original multidimensional signal, (c) real and imaginary part of MDAS are orthogonal, and (d) the magnitude envelope of a original multidimensional signal is obtained as the magnitude of its associated MDAS, which is the instantaneous amplitude of the MDAS. We have also proposed the decomposition of an image into AM-FM image model by the Fourier method (2D-FDM) and obtain explicit expression for the analytic image computation by 2D-DFT.
\section*{Acknowledgment}
The authors would like to thank JIIT Noida for permitting to carry out research at IIT Delhi.
\ifCLASSOPTIONcaptionsoff
  \newpage
\fi
\appendix[The 2D-DFT and analytic image]
In this appendix, we derive analytic image by 2D-DFT.
Let $x[m,n]$ be a real function, then the 2D discrete Fourier transform (2D-DFT) of $x[m,n]$ is defined as
\begin{equation}
X(k,l)=\frac{1}{MN}\sum_{m=0}^{M-1} \sum_{n=0}^{N-1} x[m,n] e^{-j(\frac{km}{M}+\frac{ln}{N})}  \label{2ddft_app1}
\end{equation}
 and 2D inverse discrete Fourier transform (2D-IDFT) is defined as
\begin{equation}
x[m,n]=\sum_{k=0}^{M-1} \sum_{l=0}^{N-1} X[k,l] e^{j(\frac{km}{M}+\frac{ln}{N})}.\label{2ddft_app2}
\end{equation}
In \eqref{2ddft_app1}, for odd numbers ($M=N$), there is only one real term $X[0,0]$ and $(MN-1)/2$ terms are complex conjugate of the rest $(MN-1)/2$ terms; and for even numbers ($M=N$), there are only four real terms $X[0,0],X[0,N/2],X[M/2,0],X[M/2,N/2]$ and $(MN-4)/2$ terms are complex conjugate of the rest $(MN-4)/2$ terms.
From the above discussions and using the conjugate symmetry of 2D-DFT, we obtain 2D-AS for odd numbers ($M=N$) as
\begin{multline}
z_{14}[m,n]=2\sum_{k=0}^{(M-1)/2} \sum_{l=0}^{{(N-1)}/{2}} X[k,l] e^{j(\frac{km}{M}+\frac{ln}{N})} \\
+ 2\sum_{k=(M+1)/2}^{(M-1)} \sum_{l=1}^{{(N-1)}/{2}} X[k,l] e^{j(\frac{km}{M}+\frac{ln}{N})}-X(0,0)\label{2ddft_app3}
\end{multline}
and for even numbers ($M=N$) as
\begin{multline}
z_{14}[m,n]=2\sum_{k=0}^{M/2} \sum_{l=0}^{{N}/{2}} X[k,l] e^{j(\frac{km}{M}+\frac{ln}{N})}\\+2\sum_{k=M/2+1}^{M-1} \sum_{l=1}^{{N}/{2}-1} X[k,l] e^{j(\frac{km}{M}+\frac{ln}{N})}-\\X[0,0]
-X[0,N/2]-X[M/2,0]-X[M/2,N/2],\label{2ddft_app4}
\end{multline}
where real part of AS is original signal, i.e $x[m,n]=Re\{z_{14}[m,n]\}$ and imaginary part of AS is the HT of original signal, i.e. $\hat{x}[m,n]=Im\{z_{14}[m,n]\}$.
This 2D-AS has been obtained by considering the first and fourth quadrants of 2D DFT, the second and third quadrants of 2D DFT. These 2D-AS (analytic image) computation and 2D Hilbert transform can be easily implemented with 2D-FFT algorithms, e.g. MATLAB implementation is presented in Algorithm A.
\noindent\begin{tabular}{p{0.47\textwidth}}
\hline
Algorithm A: MATLAB code for the proposed analytic image and 2D Hilbert transform computation by 2D-FFT.\\
\hline
\textbf{Case 1}: when $M$ and $N$ are even numbers.\\
\hline
\emph{X=fft2(x);[M,N]=size(x)} \% compute 2D-FFT and size.\\
\% create mask.\\
\emph{Xm=zeros(M,N);}
\emph{Xm(1:((M/2)+1),1:((N/2)+1))=2;}\\
\emph{Xm(((M/2)+2):M,2:(N/2))=2;}\% take 1st \& 4th quadrant.\\
\emph{Xm(1,1)=1; Xm(1,N/2+1)=1; Xm(M/2+1,1)=1; Xm(M/2+1,N/2+1)=1;}\\
\% mask 2D-FFT and compute inverse to obtain AS.\\
\emph{tmp=Xm.*X; z14=ifft2(tmp);}  \\
\hline
\end{tabular}
\noindent\begin{tabular}{p{0.47\textwidth}}
\hline
\textbf{Case 2}: when $M$ and $N$ are odd numbers.\\
\hline
\emph{X=fft2(x); [M,N]=size(x)} \% compute 2D-FFT and size.\\
\% create mask.\\
\emph{Xm=zeros(M,N);}
\emph{Xm(1:(M+1)/2,1:(N+1)/2)=2;} \emph{Xm(1,1)=1;}\\
\emph{Xm((M+3)/2:M,2:(N+1)/2)=2;} \% take 1st \& 4th quadrant.\\
\% mask 2D-FFT and compute inverse to obtain AS.\\
\emph{tmp=Xm.*X; z14=ifft2(tmp);}  \\
\hline
\end{tabular}

\end{document}